\RequirePackage{fix-cm}
\documentclass{svjour3}       
\smartqed  
\usepackage{rotating}
\usepackage{algorithmic}
\usepackage{graphicx}
\usepackage{textcomp}
\usepackage{xcolor, colortbl}
\usepackage{booktabs}
\usepackage{multirow}
\usepackage{hhline}
\usepackage{caption}
\usepackage{subcaption}
\usepackage{footnote}
\usepackage{amssymb}
\usepackage{url}
\usepackage{tcolorbox}
\usepackage{booktabs}
\usepackage{enumitem}
\usepackage{fixltx2e}
\usepackage{multicol}
\usepackage{balance}
\usepackage{listings}
\usepackage{xspace}
\usepackage{fontawesome5}
\usepackage[normalem]{ulem}
\usepackage[authoryear]{natbib}

\newboolean{showcomments}

\setboolean{showcomments}{true}

\newcommand{\ie}{\emph{i.e.,}\xspace}
\newcommand{\eg}{\emph{e.g.,}\xspace}

\newcommand{\etal}{\emph{et~al.}\xspace}
\newcommand{\secref}[1]{Section~\ref{#1}\xspace}
\newcommand{\figref}[1]{Fig.~\ref{#1}\xspace}

\newcommand{\tabref}[1]{Table~\ref{#1}\xspace}

\newcommand{\RQ}[1]{RQ$_{#1}$\xspace}

\newsavebox\CBox

\ifthenelse{\boolean{showcomments}}
  {\newcommand{\nb}[2]{
    \fbox{\bfseries\sffamily\scriptsize\color{red}#1}
    {\sf\small$\blacktriangleright${\color{red}\textit{#2}}$\blacktriangleleft$}
   }
   \newcommand\TBDSEC[2]{\TODO{TBD} \textcolor{#2}{\lipsum[1-#1]}}
  }
  {\newcommand{\nb}[2]{}
   
   \newcommand\TBDSEC[2]{}
  }

\newcommand\TODO[1]{\textcolor{red}{\nb{TODO}{#1}}}
\newcommand{\REV}[1]{#1\xspace}

\definecolor{gray50}{gray}{.5}
\definecolor{gray40}{gray}{.6}
\definecolor{gray30}{gray}{.7}
\definecolor{gray20}{gray}{.8}
\definecolor{gray10}{gray}{.9}
\definecolor{gray05}{gray}{.95}

\definecolor{lipsum1}{rgb}{0.00, 0.00, 1.00}
\definecolor{lipsum2}{rgb}{0.00, 0.25, 0.75}
\definecolor{lipsum3}{rgb}{0.00, 0.50, 0.50}
\definecolor{lipsum4}{rgb}{0.00, 0.75, 0.25}
\definecolor{lipsum5}{rgb}{0.25, 0.00, 0.75}
\definecolor{lipsum6}{rgb}{0.50, 1.00, 0.50}
\definecolor{lipsum7}{rgb}{0.75, 1.00, 0.25}

\newlength\Linewidth
\def\findlength{\setlength\Linewidth\linewidth
  \addtolength\Linewidth{-4\fboxrule}
  \addtolength\Linewidth{-3\fboxsep}
}
\newenvironment{examplebox}{\par\begingroup
  \setlength{\fboxsep}{5pt}\findlength
  \setbox0=\vbox\bgroup\noindent
  \hsize=0.95\linewidth
  \begin{minipage}{0.95\linewidth}\normalsize}
  {\end{minipage}\egroup
  \textcolor{gray20}{\fboxsep1.5pt\fbox
    {\fboxsep5pt\colorbox{gray05}{\normalcolor\box0}}}
  \endgroup\par\noindent
  \normalcolor\ignorespacesafterend}

\newenvironment{resultbox}{\vspace{0.2cm}\par\begingroup
  \setlength{\fboxsep}{5pt}\findlength
  \hspace{-0.5cm}
  \vspace{0.1cm}
  \setbox0=\vbox\bgroup\noindent
  \hsize=0.95\linewidth
  \begin{minipage}{0.95\linewidth}\normalsize}
  {\end{minipage}\egroup
  \textcolor{gray20}{\fboxsep1.5pt\fbox
    {\fboxsep5pt\colorbox{white}{\normalcolor\box0}}}
  \endgroup\par\noindent
  \normalcolor\ignorespacesafterend}

\begin{document}
	
	\title{Using Gameplay Videos for Detecting\\ Issues in Video Games}
	
	
	
	\author{Emanuela Guglielmi \and
		Simone Scalabrino \and
		Gabriele Bavota \and
		Rocco Oliveto
	}
	
	
	\institute{E. Guglielmi and
		S. Scalabrino and R. Oliveto \at
		STAKE Lab\\University of Molise, Italy \\
		\email \\
		{\{emanuela.guglielmi, simone.scalabrino, rocco.oliveto\}@unimol.it} \\
		\and
		Gabriele Bavota \at
		SEART @ Software Institute\\ Università della Svizzera italiana, Switzerland \\
		\email{gabriele.bavota@usi.ch}
	}

	\date{Received: date / Accepted: date}

	\maketitle
	
	\newcommand{\approach}{GELID\xspace}
    \newcommand{\approachlong}{GamEpLay Issue Detector\xspace}
	
	\begin{abstract}
		\textit{Context.} The game industry is increasingly growing in recent years. Every day, millions of people play video games, not only as a hobby, but also for professional competitions (\eg e-sports or speed-running) or for making business by entertaining others (\eg streamers). The latter daily produce a large amount of gameplay videos in which they also comment live what they experience. But no software and, thus, no video game is perfect: Streamers may encounter several problems (such as bugs, glitches, or performance issues) while they play. Also, it is unlikely that they explicitly report such issues to developers. The identified problems may negatively impact the user's gaming experience and, in turn, can harm the reputation of the game and of the producer.
        \textit{Objective.} In this paper, we propose and empirically evaluate \approach, an approach for automatically extracting relevant information from gameplay videos by (i) identifying video segments in which streamers experienced anomalies; (ii) categorizing them based on their type (\eg logic or presentation); clustering them based on (iii) the context in which appear (\eg level or game area) and (iv) on the specific issue type (\eg game crashes).
        \textit{Method.} We manually defined a training set for step 2 of \approach (categorization) and a test set for validating in isolation the four components of \approach. In total, we manually segmented, labeled, and clustered 170 videos related to 3 video games, defining a dataset containing 604 segments.
        \textit{Results.} While in steps 1 (segmentation) and 4 (specific issue clustering) \approach achieves satisfactory results, it shows limitations on step 3 (game context clustering) and, above all, step 2 (categorization).

    \keywords{video games, gameplay videos, mining software repositories}
	\end{abstract}
	
	\section{Introduction}
\label{sec:introduction}
Video games are becoming an increasingly important form of expression in Today's culture. Their sociological, economic, and technological impact is well recognized in the literature \citep{jones2008meaning} and their wide diffusion, particularly among the younger generations, has contributed to the growth of the gaming industry in several directions. Playing video games is progressively becoming a work for many: Some play for professional competitions (\eg in e-sports or speed-running), while others play to entertain others (\eg streamers) especially on dedicated platforms such as Twitch\footnote{\url{https://twitch.tv}}.
Besides all challenges that are common to software systems, developing and maintaining video games pose additional difficulties related to complex graphical user interfaces, performance requirements, and higher testing complexity. Concerning the latter point, games tend to have a large number of states that can be reached through different choices made by the player. In such a context, writing automated tests is far from trivial due to the need for an ``intelligent'' interaction triggering the states exploration. Even assuming such ability to explore the game space, determining what the correct behavior is in a specific state usually requires human assessment, with the exception of bugs causing the game to crash. Finally, additional complexity is brought by the non-determinism that occurs in games because of multi-threading, distributed computing, artificial intelligence and randomness injected to increase the difficulty of the game \citep{murphy2014cowboys}. 

Because of the few automated approaches available for quality control in video game development \citep{santos2018computer}, many games are released with unknown problems that are revealed only once customers start playing \citep{truelove2021we}. Since many streamers daily publish hours of gameplay videos, it is very likely that some of them experience such issues and leave traces of them in the uploaded videos. For example, a gameplay video on the game Cyberpunk 2077\footnote{\url{https://youtu.be/ybvXzSLy9Ew?t=1448}} shows that the game crashes as soon as the player performs a specific action.
The large amount of publicly available gameplay videos, therefore, might be a goldmine of information for developers. Indeed, such videos not only contain information about which kinds of issues affect a video game, but they also provide examples of interactions that led to the issue in the first place, allowing its reproduction. 
In their seminal work on this topic, \citet{lin2019identifying} defined an approach able to automatically identify videos containing bug reports. However, such an approach mostly relies on the video metadata (\eg its length) and it is not able to pinpoint the specific parts of the video in which the bug is reported. This makes it unsuitable as a reporting tool for game developers, especially when long videos, which are not uncommon, are spot as bug-reporting.

In this paper, we introduce \approach (\approachlong), an automated approach that aims at complementing the approach by \citet{lin2019identifying} by (i) automatically extracting meaningful segments of gameplay videos in which streamers report issues, and (ii) hierarchically organize them. 
Given some gameplay videos as input, \approach (i) partitions them into meaningful segments that might contain bug reports, (ii) automatically distinguishes informative segments from non-informative ones by also determining the type of reported issue (\eg bug, performance-related), (iii) groups them based on the ``context'' in which they appear (\ie whether the issue manifests itself in a specific game area), and (iv) clusters fragments related to the same specific issue (\eg the game crashes when a specific item is collected). 

We evaluate the four components of \approach in isolation, to understand to what extent it is possible to achieve the single goals we set, as we planned in our registered report presented at MSR 2022 \citep{guglielmi2022towards}. We first extract training data for the machine learning model we use to categorize segments (step 2 of \approach). To this end, we used the approach by \citet{lin2019identifying} to identify candidate videos from which we can manually label segments in which the streamer is reporting an issue. Then, we ran \approach on a set of real gameplay videos and validate its components. First, we manually determine to what extent the extracted segments are usable, by annotating their \textit{interpretability} (\ie they can be used as standalone videos) and \textit{atomicity} (\ie they can not be further split). Second, we validate the categorization capabilities of \approach, both when trying to distinguish non-informative segments from informative ones (binary classification) and when trying to pinpoint the specific issue among \textit{logic}, \textit{performance}, \textit{presentation}, \textit{balance}, and \textit{non-informative} through a multi-class classifier. To this and, we compute typical metrics used to evaluate ML models (\ie accuracy and AUC).
Finally, we evaluate to what extent the clusters identified in terms of context and specific issues are similar to the manually determined ones using the MoJoFM metric \citep{wen2004effectiveness}.

The remainder of this paper is organized as follows. In \secref{sec:related} we present the background needed for understanding the paper and some related work, based on which we also define the four specific categories of issues that \approach will identify. In \secref{sec:approach}, we present \approach and its four components in details. In \secref{sec:design} we describe the empirical study design, while in \secref{sec:results} we report the obtained results. In \secref{sec:discussion} we discuss the results, while in \secref{sec:threats} we report the threats to validity. \secref{sec:conclusions} concludes the paper.

	\section{Background and Related Work}
\label{sec:related}

The large efforts that game developers invest in the game development process do not always allow them to discover or fix all the bugs in a game before releasing it to the market. Several works have focused the attention on the quality assurance of video games analyzing the differences between traditional software development and video games development \citep{murphy2014cowboys, santos2018computer}. Many studios employ discussion forums or specific features in their games for gamers to report bugs (\eg Steam Community). Previous work shows that 80\% of the Steam games release urgent updates to fix issues such as feature malfunctions or game crashes \citep{lin2017studying}. The large amount of gameplay videos continuously produced and publicly released by many gamers on platforms such as Twitch and YouTube could be helpful to developers: Sometimes, gamers indirectly report issues while they play. Since \approach aims to support video game developers by extracting information from gameplay videos, the discussion focuses mainly on approaches aimed at extracting and manipulating gameplay videos for different purposes. In addition, since our approach aims to automatically categorize video segments, we also discuss existing taxonomies of video game topics that we use as a starting point for defining our categories.

\subsection{Mining of Gameplay Videos}
Some works targeted the automated generation of a comprehensive description of what happens in gameplay videos (\ie game commentary). Examples of these works are the framework by \citet{guzdial2018towards} and the approach presented by \citet{li2019end} modeling the generation of commentaries as a sequence-to-sequence problem, converting video clips to commentary. On the same line of research, an approach to generate automatic comments for videos by using deep convolutional neural networks was presented by \citet{shah2019automated}.
\citet{lewis2010went} described the gameplay videos as ``a rich resource.''
The main goal of \approach is to detect issues in gameplay videos. To the best of our knowledge, the only work aimed at achieving a similar goal is the one by \citet{lin2019identifying}. The authors conducted an in-depth study of gameplay videos posted by players on the Steam platform aiming at automatically identifying the ones that report bugs. They observe that na\"{\i}ve approaches based on keywords matching are inaccurate. Therefore, they propose an approach that uses a Random Forest classifier \citep{ho1995random} to categorize gameplay videos based on their probability of reporting a bug.
\citet{lin2019identifying} rely on Steam\footnote{\url{https://steamcommunity.com/}} to find videos related to specific games. While Steam is mainly a marketplace for video games, it also allows users to interact with each other and share videos. On a daily basis, for 21.4\% of the games on Steam, users share 50 game videos, and a median of 13 hours of video runtime \citet{lin2019identifying}. 
Still, their approach works at video-level, and manually watching long gameplay videos classified as buggy still requires a considerable manual effort since a whole video can even last several hours. Also, they only distinguish bug-reporting videos from non-bug-reporting ones, without a more specific classification regarding the type of issue reported (\eg glitch or logic bug).
We fill this gap and further aid developers by classifying the video segment according to the type of problem encountered, and by trying to classify video segments (\ie parts of videos) instead of whole videos. To achieve this goal, \approach augments the provided information, by including also (i) the type of issue found, (ii) the context (\ie area of the game) in which it occurred, and (iii) other segments in which the same issue was reported (possibly from different videos).

\subsection{Taxonomies of Video Game Issues}
\label{sec:related:taxonomy}
Video games can suffer from a vast variety of problems. Lin \etal \citep{lin2019identifying} do not distinguish among the types of issues reported in the videos identified as ``bug reporting'', while this is one of our goals. 

To determine meaningful categories in which it is worth categorizing video segments, we rely on a recent taxonomy of issues in video games introduced by \citep{truelove2021we} (which extends the one by  \citep{lewis2010went}). In their taxonomy, the authors reports 20 different kinds of issues. 

\noindent We use such a taxonomy as a base to define the labels we want to assign to the video segments. However, all such labels might be counterproductive since it is likely to observe a long-tail distribution (\ie a few types of issues appear in most of the video fragments, while several other issues are quite rare or do not even appear). Therefore, starting from such a taxonomy, we define macro-categories by clustering similar fine-grained categories. We identified four labels, as reported in \tabref{tab:taxonomy}: \textit{Logic}, \textit{Presentation}, \textit{Balance}, and \textit{Performance}. 

\begin{table}
  \centering
  \caption{Mapping between types of issues identified by \approach and categories from the taxonomy by Truelove \etal \citep{truelove2021we}.}
  \label{tab:taxonomy}%
  \resizebox{0.9\linewidth}{!}{
    \begin{tabular}{l p{3cm} l}
    \toprule
    \textbf{Issue Type} & \textbf{Description} & \textbf{Categories \citep{truelove2021we}}\\
    \midrule
    \multirow{11}{*}{\textbf{Logic}} & 
    \multirow[t]{11}{*}{
    \parbox[t]{3cm}{Issues related to the game logic, regardless of how information is presented to the player.
    }} & 
        Object Persistence \\
    & & Collision of Objects\\
    & & Inter. btw. Obj. Prop.\\
    & & Position of Object\\
    & & Context State\\
    & & Crash\\
    & & Event Occurrence\\
    & & Interrupted Event\\
    & & Triggered Event\\
    & & Action\\
    & & Value\\
    \midrule
    
    \multirow{6}{*}{\textbf{Presentation}} & 
    \multirow[t]{6}{*}{
    \parbox[t]{3cm}{Issues related to the game interface (graphical- or audio- related).
    }} & 
        Game Graphics\\
    & & Information\\
    & & Bounds\\
    & & Camera\\
    & & Audio\\
    & & User Interface\\ 
    \midrule
    
    \multirow{2}{*}{\textbf{Balance}} & 
    \multirow[t]{2}{*}{
    \parbox[t]{3cm}{
    Detrimental aspects in terms of ``fun''.
    }} & 
        Artificial Intelligence\\
    & & Exploit\\
    \midrule
    
    \multirow{2}{*}{\textbf{Performance}} & Performance-related issues (\eg FPS drops). & Implem. Response\\
    \bottomrule
    \end{tabular}%
    }
\end{table}%

 
\section{\approach} \label{sec:approach}

\approach takes as input a set of gameplay videos related to a specific video game and returns a hierarchy of segments of gameplay videos organized on three levels: (i) context (\eg level or game area), (ii) issue type (\eg bug or glitch), and (iii) specific issue (\eg game crashes when talking to a specific non-player character).

\begin{figure} 
	\centering\includegraphics[width=\linewidth]{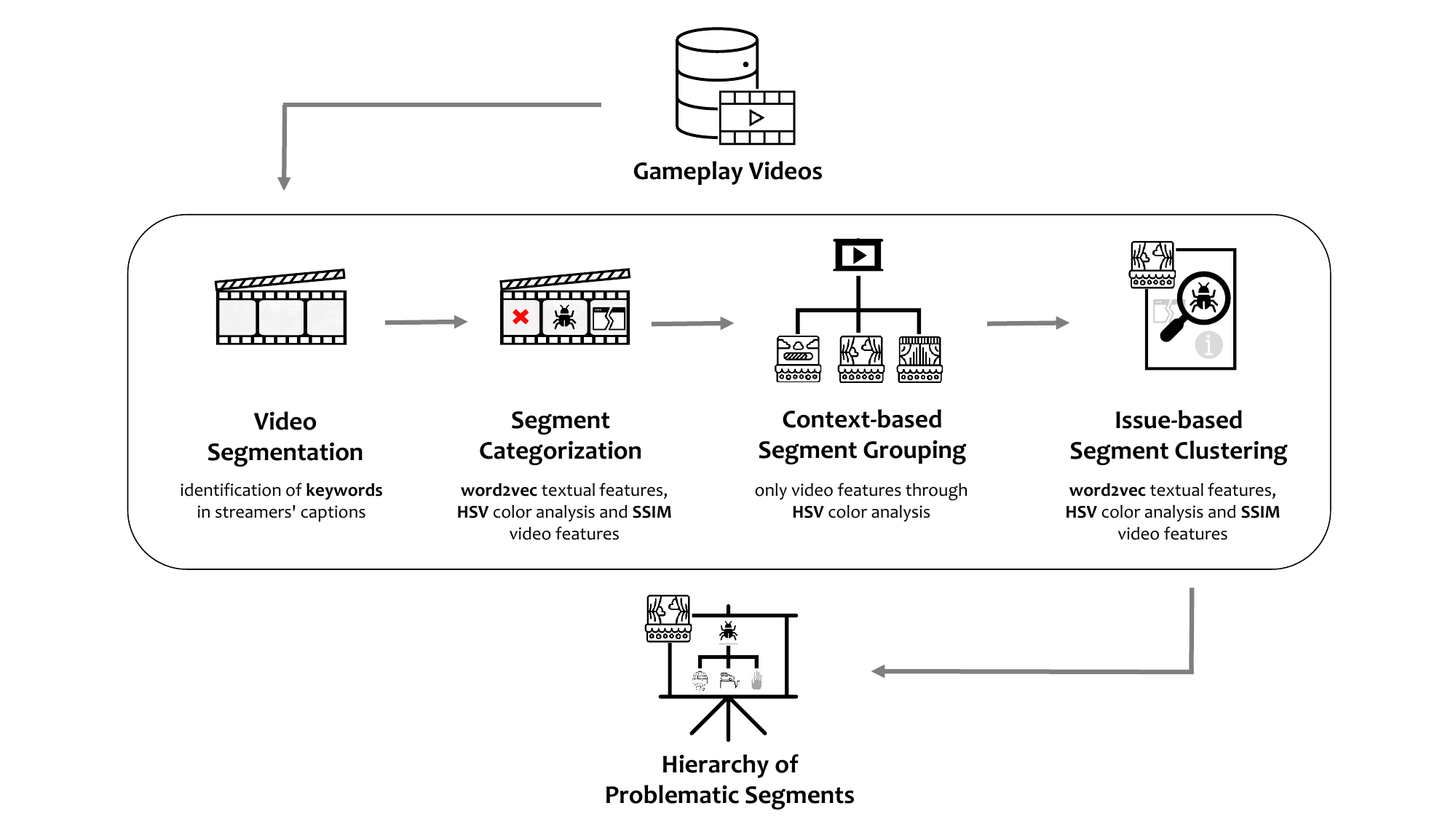}
	\captionsetup{justification=centering}
	\caption{The workflow of \approach.}
	\label{fig:workflow}
\end{figure}

\figref{fig:workflow} shows an overview of the \approach workflow. We describe below in more detail the main steps of \approach.

\subsection{Video Segmentation}
\label{sec:videosegmentation}
The first step of \approach consists in partitioning the video into meaningful segments that can be later analyzed as standalone shorter videos. 
In the computer vision literature, a similar problem is referred to as ``shot transitions detection'' \cite{souek2020transnet}. The aim is to detect sudden changes in the video content. An example of approaches defined to solve such a problem is the one introduced in \citep{tang2018fast}.
Video-related information, however, might not be sufficient to find cuts in gameplay contents. \REV{In the context of video segmentation, relying only on scene changes to identify meaningful segments may not be sufficient. Scene changes may be due to various minor factors, \eg rapid zoom into the viewfinder of a weapon and then back to the general framing of the scene. Such situations do not provide significant information for identifying potential issues. Furthermore, in some contexts, scene changes may not be evident, leading to the creation of very large segments that are difficult to analyse. Let us consider, for example, the gameplay video available at \url{https://www.youtube.com/watch?v=_kQIJ2Omy9w}: From 14:10 to 15:56 there is no shot transition, even though various separate events and actions occur.} Moreover, for example, if the game crashes and a shot transition detection approach is used to cut the video, the second in which the crash happens would probably be selected for segmentation. The streamer, however, might need a few seconds to react to such an event by commenting what happened providing useful information for the game developers. Thus, by using shot transitions as cut points, the spoken content related to the issue might be erroneously put in the subsequent segment. To solve this problem, we decided to mainly rely on the spoken content to decide the cut points in the video: The core idea is to get the points in which each subtitle entry (\ie units of text shown on the screen) begins and ends, slightly shifted by \REV{$t$} seconds (where \REV{$t$} is a parameter of the approach) to take into account the reaction time of the streamer, and thus consider the video in-between as a segment. As for the shifting operation, given a subtitle entry that starts at second $s$ and ends at second $s+d$ (where $d$ is the duration of the subtitle entry), our approach will extract the video segment between $\max(s-t, 0)$ and $\min(s+d+t, \mathit{video\ length})$. For example, consider the case where we set \REV{$t = 5$} and we detect a subtitle entry that starts at 13:45 (mm:ss) and lasts 3 seconds. Our approach will cut the video between 13:40 and 13:52. We report in \secref{sec:design} how we tune the \REV{$t$} parameter.

As a result, our segmentation approach will implicitly discard some parts of the input video (\ie the ones in which the streamer is not speaking) and it might put some parts of the video in many segments when \REV{$t > 0$} (\eg for contiguous subtitle entries). Also, it is worth noting that using this strategy might result in a very high number of extracted segments for each video since subtitle entries generally include only parts of a sentence: In subtitles, a given sentence is broken into several entries to allow the watcher to comfortably read each of them. To preliminarily exclude segments that most likely do not contain any piece of useful information and, thus, to reduce the effort for the next step, we use a keyword-matching approach. If at least a relevant keyword is found in the subtitle entry related to a given segment, we consider the segment, while we exclude it otherwise. 

To define the list of keywords, we relied on (i) the 12,122 change notes of video games used by \citet{truelove2021we} to define the taxonomy of the most frequently encountered problems in video games and (ii) the 996 titles and descriptions of the gamplay videos in the dataset defined by \citet{lin2019identifying}. One of the authors manually extracted, from each instance, a first set of keywords (also composed by more than a word) which were related to issues in video games (\eg ``glitch'' or ``bug''). As a result of this process, 161 basic keywords were identified the file containing the selected keywords is reported in the replication pacakage \citep{replicationpackage}. 
From such keywords, we automatically generated new semantically equivalent keywords to have a broader dictionary.
To do that, we first tokenized the keywords and automatically tagged the Part-of-Speech (PoS) by using the spaCy Python package \citep{spacy}. Then, for each token with its PoS tag, we used both WordNet \citep{miller1995wordnet} and SEWordSim \citep{tian2014sewordsim} to generate both general-purpose and domain-specific synonyms of each word.
At this point, for each keyword composed by the sequence of words $\langle w_1, \dots, w_n \rangle$, we combined all the synonyms of each word and generated the new set of candidate keywords by using the Cartesian product: $\{ \mathit{syn}(w_1) \times \dots \times \mathit{syn}(w_n) \}$. For example, given the initial keyword ``lag'', we generated the candidate alternative keywords ``stuttering'', ``FPS drop''. From the initial 161 identified keywords, we obtained a total of 207 candidate keywords.
Then, two of the authors independently validated the new keywords to discard the ones that were not related to issues in video games. In case of disagreement, they discussed to reach consensus. In the end, we added 96 new keywords, while 111 were discarded. \REV{In our analysis, we assessed the inter-rater reliability between the annotators involved in identifying keywords by calculating Cohen's Kappa coefficient. The obtained results indicate an agreement level of $k = 0.74$. The coefficient value of 0.74 indicates a good level of agreement between the annotators in terms of identifying the keywords. }For example, the keyword ``crash'' generated from ``break up'' was discarded. Thus, our final list of keywords is composed of 257 keywords which can be mapped to our replication package \citep{replicationpackage}.

\subsection{Segment Categorization} 
\label{sec:videocategorization}
In this second step, \approach aims at categorizing segments based on their content. \approach considers five labels: One for \textit{non-informative} segments (\ie the ones not reporting issues), and four for \textit{informative} segments (\ie the ones reported in \tabref{tab:taxonomy}). Non-informative segments are discarded and not considered in the next steps.

Previous work successfully used machine-learning to solve similar classification problems in the context of mobile app reviews \citep{chen2014ar, scalabrino2017listening}. Such approaches mainly rely on textual features. In our context, we can extract information that could also help to correctly classify segments from video analysis. For example, segments without video might be more likely to be \textit{non-informative}, even if a reader comment is present. Therefore, we include in \approach also video-based features. 
More specifically, we extract five sets of features: Three of them only based on the subtitles (\ie what the streamer says), one of them based on the video (\ie what happens in the game), and one of them including the best set of textual features and the set of video-based features.

\textbf{Textual Features.}
As for the textual features, we consider Bag of Words (BoW) \citep{zhang2010understanding}, doc2vec (d2v) \citep{karvelis2018topic} and word2vec (w2v) \citep{rong2014word2vec}. BoW consists in detecting the dictionary of the training set and using each word of the dictionary as a feature. The value of each feature for a given instance corresponds to the number of times the related word appears in an instance. The number of features directly depends on the training set. In our case, given the training set described in \secref{sec:datasets}, we extracted 2,253 features.
The d2v model \citep{karvelis2018topic} allows to automatically extract a vector of features for an entire instance (document). Such a model allows to automatically represent a document (sequence of words) as a vector. Specifically, we represent each subtitle string \REV{for each identified segment} as a vector composed of 40 features \REV{since this is the default number of features extracted by such a model \citep{karvelis2018topic}}.
Finally, the w2v model \citep{rong2014word2vec} allows to represent a single word as a set of features. Thus, differently from doc2vec, it does not directly work at document-level. To define the features based on w2v, given all the words in a given instance, we extract the vectors through the w2v model and we compute the average of each feature. In this case, we represent each word as a vector of 300 features, \REV{again, because the w2v model extracts by default such a number of features \citep{rong2014word2vec}}.

\textbf{Video-based Features.}
With video-based features, instead, we mainly wanted to represent to what extent the video contains unexpected frames that could possibly be related to issues. To this aim, given each pair of subsequent frames $f_i$ and $f_{i+1}$: (i) we compute their structural similarity through SSIM \citep{ssim}, \ie $s_i = \mathit{SSIM}(f_i, f_{i+1})$; (ii) we extract their HSV histograms using the HISTCMP CORREL function of OpenCV \cite{opencv}, thus obtaining $h(f_i)$ and $h(f_{i+1})$; (iii) we then compute their Pearson correlation coefficient $\mathit{hsv}_i = \mathit{cor}(h(f_i), h(f_{i+1}))$. 
We use SSIM instead of other image similarity measures because it has been shown that such a metric best captures the similarity of images as perceived by humans \citep{ssim}. Since such a metric ignores colors but considers, by default, a black-and-white version of the image, we also use HSV histograms to detect differences in the colors.
Finally, given the vectors of values $\mathit{hsv}$ and $\mathit{s}$ for all the frames between 0 and $n$ (number of frames in the video), we aggregate their values and define 12 video based features by computing the mean, median, minimum, maximum, first quartile, and third quartile of both of them. Such features allow us to inform the model about the distribution of such vectors. For example, let us imagine that the game crashed: the frame $f_i$ before the crash is very similar to the previous ones, while the next frame, $f_{i+1}$ is different from $f_i$. As a result, both $\mathit{hsv}_i$ and $s_i$ will be very high. Two of our features (\ie the max of both the vectors) will reflect this information.

Given a training set of labeled video segments, we extract the features and train a ML classifier. Given an input (unknown) video segment, we extract the same features used to train the model, given the resulting vector as input to the trained ML model, and obtain the predicted label. We describe in \secref{sec:design} how we built the training set and how we select the best ML algorithm for this task among Random Forest \citep{ho1995random}, Logistic Regression, SMO \citep{hearst1998support}, Multilayer Perceptron \citep{ramchoun2016multilayer} and IBk\citep{choudhury2015comparative}.

\subsection{Context-based Segment Grouping} 
\label{sec:contextgrouping}
After having collected and categorized segments that contain anomalies (\ie the ones classified as \textit{informative}, \ie as \textit{logic}, \textit{presentation}, \textit{performance}, or \textit{balance}), we group them according to their \textit{context}.
With ``context'' we refer to the part of the game (\eg a specific game level or area) in which the anomaly occurred. This may be helpful to provide the videos to the team in charge of the development of that specific part of the game.
Such a step is important for two reasons: (i) Developers analyzing hundreds of videos related to a specific game may experience information overload and this, in turn, would reduce the effectiveness of the video segments filtering step; (ii) Knowing the context in which more anomalies occur allows the developer to identify where attention needs to be focused to improve the gaming experience.

To achieve this goal, we rely on video information: The assumption is that videos with similar frames regard, most likely, the same context.
First, we extract the key frames from each segment by using the Video-kf Python package \citep{videokf}. Then, we define a summary frame of the whole segment by computing a pixel-by-pixel average of the previously identified key frames. Such a frame will roughly represent the content of the segment and, ideally, it can allow to visually represent the game area.
We use a clustering algorithm to group summary frames (and, thus, the associated segments). More specifically, given a distance function between two images (summary frames, in our case), we define a distance matrix which contains the distances between each couple of summary frames and use it to cluster them.

We test two similarity metrics (which are also used for computing the video-based features in the previous step): Structural similarity (SSIM) \cite{ssim}, computed on each pair of summary frames, and the correlation between the HSV histograms extracted from each pair of summary frames. Note that both of them are \textit{similarity} metrics, while clustering algorithms require to indicate the \textit{distances} between instances. Since both of them are bounded in the range $[0, 1]$, we simply transform them in distance metrics by computing $1 - s$ (where $s$ is the value of the similarity metric).

Since the number of scenes is not necessarily known \emph{a priori}, we use a non-parametric clustering technique. We describe in \secref{sec:design} how we select the best clustering algorithm between the two we tested, \ie DBSCAN \citep{ester1996density} and OPTICS \citep{ankerst1999optics}, and the best distance metric between SSIM and HSV histogram correlation.

\subsection{Issue-based Segment Clustering}
\label{sec:issuegrouping}
A set of video segments of the same kind (\eg bugs) and reported in the same context might still be hard to manually analyze for developers. For example, if 100 segments report bugs for a given level, developers need to manually analyze all of them. It might be the case, however, that most of them report the same specific bug (\eg a game object disappears). To reduce the effort required to analyze such information, we cluster segments reporting the same specific issue. This would allow developers to analyze a single segment for each cluster to have an overview of the problems affecting the specific area of the game.

To achieve this goal, we represent the instances (\ie video segments) by using both textual and image-based features and, as in the previous step, we use non-parametric clustering to create homogeneous groups.
Textual features can help grasping the broad context (\eg objects disappearing or anomalous dialogues). Image-based features can help finding visually similar problems (\eg in the case of glitches). To this aim, we represent each instance (video segment) using the set of features from the categorization step that allows to obtain the best results for that task (as we report in \secref{sec:results}). Differently from the previous step, indeed, we do not pre-compute the distance matrix. This allows us to test this task not only with DBSCAN \citep{ester1996density} and OPTICS \citep{ankerst1999optics}, but also with Mean Shift \citep{fukunaga1975estimation}, which, differently from the previously-mentioned algorithms, does not allow to directly use a distance matrix. Also in this case, we describe in \secref{sec:design} how we select the best clustering algorithm among them.

	\section{Empirical Study Design}
\label{sec:design}

The goal of our study is to evaluate the effectiveness of the four steps of \approach, \ie (i) extraction of meaningful video segments from gameplay videos (ii) accuracy in categorizing extracted video segments, (iii) capability of clustering video segments about the same gameplay area, and (iv) ability to correctly cluster segments reporting the same specific issue. The context of the study consists of a total of 275 gameplay videos.

Our study is steered by the following research questions (RQs).\begin{resultbox}
 \textbf{\RQ{1}}: \textit{How meaningful are the gameplay video segments extracted by \approach?}
\end{resultbox}
The first RQ aims at evaluating the quality of the segments extracted by \approach from gameplay videos in terms of their \textit{interpretability} and \textit{atomicity}. It aims at evaluating the ``video segmentation'' step described in \secref{sec:videosegmentation}.

\begin{resultbox}
 \textbf{\RQ{2}}: \textit{To what extent is \approach able to categorize gameplay video segments?}
\end{resultbox}
With this second RQ we want to understand which features and which classification algorithm allow to train the best model for categorizing gameplay video segments both in two classes (\textit{informative} and \textit{non-informative}, like previous work \cite{lin2019identifying}) and five classes (\textit{logic}, \textit{presentation}, \textit{performance}, \textit{balance}, and \textit{non-informative}). We also want to understand to what extent the best models for the two categorization problems would allow to achieve useful results in practice. \RQ{2} evaluates the ``segment categorization'' step described in \secref{sec:videocategorization}.

\begin{resultbox}
 \textbf{\RQ{3}}: \textit{What is the effectiveness of \approach in grouping gameplay video segments by context?}
\end{resultbox}
In the third RQ, we aim to understand what the best clustering algorithm is for grouping segments based on the game context, and how effective such an algorithm is in absolute terms. This RQ evaluates the clustering step described in \secref{sec:contextgrouping}.

\begin{resultbox}
 \textbf{\RQ{4}}: \textit{What is the effectiveness of \approach in clustering gameplay video segments based on the specific issue?}
\end{resultbox}
Similarly to \RQ{3}, \RQ{4} aims at understanding which features and clustering algorithm allow to achieve the best results for clustering segments based on the specific issue, and how effective such an algorithm is in absolute terms. This RQ evaluates the clustering step described in \secref{sec:issuegrouping}.


\subsection{Context Selection}
\label{sec:datasets}
To the best of our knowledge, there are no large-scale, publicly available databases of gameplay videos that provide meaningful information on the classification of problems in video games through subtitle analysis. To answer our RQs and validate the defined approach, we rely on gameplay videos from YouTube. While other platforms, even more video game-oriented, could be used (\eg Twitch), YouTube provides APIs for searching videos of interest and it also allows to download videos including subtitles, which are required by \approach. While subtitles can be automatically generated when the video lacks them, the results could be noisy and, in this phase, we evaluate \approach assuming high-quality input data. \REV{In our study, we collect three datasets, and the criteria used to search for gameplay videos of interest depend on the dataset at hand (explicited in the subsections below).}

The first dataset is composed by video segments, and we use it used for training the supervised model used in step 2 of \approach (\ie segment categorization). We also use this dataset to select the best model for answering \RQ{2}. The second one is composed by complete videos, and we use it for evaluating the single components of \approach and answer \RQ{1-4}. The third one is a smaller dataset used to evaluate the parameters to be used in the different feature extraction and machine-learning techniques. We publicly release all datasets in our replication package \cite{replicationpackage}.

\subsubsection{Training Data}
Our goal is to build a training set of labeled segments containing at least 1,000 instances and covering all the issue types \approach is able to identify.
To select videos possibly useful to build our training set, we used the YouTube Search APIs\footnote{\url{https://developers.google.com/youtube/v3}}. Specifically, we ran a query using the same keywords used by \citet{lin2019identifying}, \ie ``bug'', ``hack'', ``glitch'', ``hacker'', ``cheat'', and ``cheater''. For each keyword, we retrieved a list of videos matching it. We also added a filter to exclude videos without subtitles or with subtitles in languages different from English since \approach relies on NLP-based features computed on them.
Some YouTube videos have manually-defined subtitles, while others have automatically generated ones. We include both of them. Indeed, while it is possible that the second category contains errors, this risk also exists in manually generated ones. Also, the quality of the subtitles generated by YouTube is generally quite high for the English language. As a result, we obtained 3,540 videos. Since some videos were present in more than a list (\ie they matched different keywords), we removed duplicates and obtained 3,196 videos. \REV{We report in \tabref{tab:trainingvideoselection}, for each keyword, the number of videos retrieved and filtered, along with the number of extracted segments. Note that the number of segments might be lower than the number of filtered videos because a video might not contain valid keywords in the subtitles even though it contains them in other metadata, such as the title.}

\begin{table}
 \caption{Number of videos retrieved for each keyword.}
 \label{tab:trainingvideoselection}
 \centering
 \begin{tabular}{lrrr}
  \toprule
  \textbf{Keyword}  & \textbf{\#Videos Retrieved} & \textbf{\#Filtered Videos} & \textbf{\#Segments}\\
  \midrule
  \textit{bug}      &  594  & 514 & 691     \\
  \textit{glitch}   &  509  & 487 & 282     \\
  \textit{hack}     &  514  & 155 & 64     \\
  \textit{hacker}   &  502  & 115 & 66     \\
  \textit{cheat}    &  528  & 145 & 112     \\
  \textit{cheater}  &  549  & 118 & 40     \\
  \bottomrule
 \end{tabular}

\end{table}

Our premise is that several gameplay videos report issues. However, issue-reporting videos represent a minority of the entire gameplay videos population (thus the relevance of our research). Therefore, to support the construction of the dataset containing training data for the categorization step, we relied on the approach defined by \citet{lin2019identifying} and consider only videos identified as issue-reporting. 
Specifically, we re-implemented their approach (since it is not publicly available) and, for each video retrieved as previously described, we ran the approach and discarded the videos classified as non-issue-reporting. As a result, we kept 1,534 videos. We shuffled such videos and manually analyzed them one by one to extract and label segments. 
One of the authors manually split each video into meaningful segments, and two of the authors manually labeled each segment as \textbf{logic}, \textbf{presentation}, \textbf{balance}, \textbf{performance}, or \textbf{non-informative} (when the segment does not report any issue). \REV{Specifically, in order to manually split the video into segments, one of the authors carefully watched each gameplay video, covering its entire duration. During this process, the author noted down the specific starting and ending times (in seconds) for each segment that they identified within the video. The identification of significant segments was guided by a specific criterion based on the classification outlined in Table 1, which can be found in \secref{sec:related:taxonomy}.
With the phrase ``meaningful segments'' we mean video segments that can be analyzed independently as shorter videos and contain enough information that can help achieve the objectives of GELID. To determine whether a segment is ``meaningful,'' as we report later, we use the principles of \textit{interpretability} (to what extent humans can get information from the segment) and \textit{atomicity} (to what extent the segment contain only the information related to a single issue).} At this stage, we discarded segments reporting more than an issue at a time. 
Given the large quantity of videos available compared to the target number of segments we had in mind, we decided to make sure that the training set was diverse in terms of video games considered. Thus, if we noticed that a video game was already taken into account in several videos previously analyzed, we avoided to analyze more videos of it. 
In total, we manually analyzed 170 gameplay videos, totaling about 17 hours of gameplay. As a result, we identified and labeled 1,255 video segments.

Specifically, we obtained 693 non-informative video segments ($\sim$55.2\%), 305 video segments reporting presentation-related problems ($\sim$24.3\%), 169 video segments reporting logic problems ($\sim$13.5\%), 47 video segments with balance problems ($\sim$3.7\%), and 41 video segments highlighting performance problems ($\sim$3.3\%). Given the nature of the problem at hand, as we expected, the dataset is imbalanced, with a great majority of segments being non-informative and a very small percentage of them reporting balance- and performance-related issues.

\subsubsection{Components Validation Data (Test Set)}
To select videos on which we validate the single components of \approach, we focused on a small set of video games. We did this because the third and fourth steps of \approach are reasonable only when segments from the same video game are considered. To select the video games to use, we rely on the information available on Steam, one of the largest video game marketplaces \citep{toy2018large}. \REV{Based on information obtained from Steam} we select three video games that are both popular (\eg for which many gameplay videos exist) and that had several reported issues (\eg for which \approach gives the best advantage). More specifically, we select video games with many downloads and low review scores. To do this, we first retrieved the list of the top 100 most downloaded games on Steam, as reported in \tabref{tab:gamessteam}. Then, we excluded the games with \textit{very positive} or better reviews (\ie we kept the ones with ``mostly positive'' reviews or lower). 
We preliminarily analyzed a random sample of 10 gameplay videos for each video game after this filter using the YouTube search feature. If we found no gameplay videos reporting issues, we discarded the video game. Then, for all the remaining video games, we used the YouTube Search APIs to search for ``\textit{video-game-name} gameplay video''. We applied filters to select only videos with English subtitles (either manually added or automatically generated) and with medium (4-20min) and long (+20min) duration, with the aim of excluding non-informative videos representing game trailers or identifying a compilation of issues (which, instead, were useful to build the training set). Finally, we selected the three video games with the highest number of gameplay videos retrieved, \ie \textit{Conan Exiles} \citep{conan}, \textit{DayZ} \citep{dayz} and \textit{New World} \citep{newworld}. In total, we obtained 80 gameplay videos, totaling about 45 hours of gameplay.

Since manually splitting the entire videos would have been very demanding, we decided to partially rely on the first step of \approach. More specifically, we identified in the subtitles the keywords selected for the segmentation step. Then, one of the authors manually segmented the video near those points to select a first set of possibly relevant segments, and two of the authors independently manually categorized and clustered them both based on the context and on the specific issue (only for informative videos). The two annotators discussed conflicts to reach consensus.
In total, we identified 604 video segments, distributed as depicted in \tabref{tab:testsetdistribution}. It is worth noting that we were able to identify only a few balance-related segments (4 in total, with DayZ having none of them).

\begin{table}
 \newcommand{\Lo}{\faIcon[regular]{bug}}
 \newcommand{\PR}{\faIcon[regular]{ghost}}
 \newcommand{\Pf}{\faIcon[regular]{clock}}
 \newcommand{\Bl}{\faIcon[regular]{balance-scale-right}}
 \newcommand{\NI}{\faIcon[regular]{trash}}

 \caption{Distribution of issue types (logic \Lo, presentation \PR, performance \Pf, balance \Bl, and non-informative \NI) for each video game considered in the test set.}
 \label{tab:testsetdistribution}
 
 \centering 
 \begin{tabular}{l | r r r r r | r}
 \toprule
 \textbf{Video Game} & \Lo   & \PR   & \Pf    & \Bl    & \NI   & \textbf{Total} \\
 \midrule
 Conan Exiles        & 37    & 109   & 10     & 1      & 157   & 314 \\
 DayZ                & 7     & 67    & 16     & 0      & 90    & 180 \\
 New World           & 2     & 44    & 6      & 3      & 55    & 110 \\
 \midrule
 Total               & 46    & 220   & 32     & 4      & 302   & 604 \\
 \bottomrule
 \end{tabular}

\end{table}

\begin{table*}[t]
 \newcommand{\OverwhelminglyPositive}{\faIcon[regular]{check}\faIcon[regular]{check}\faIcon[regular]{check}}
 \newcommand{\VeryPositive}{\faIcon[regular]{check}\faIcon[regular]{check}}
 \newcommand{\MostlyPositive}{\faIcon[regular]{check}}
 \newcommand{\Mixed}{$\mathbf{\thicksim}$}
 \newcommand{\MostlyNegative}{\faIcon[regular]{times}}
 \centering
 \caption{Top 100 most popular games on Steam and related summary review scores (``Overwhelmingly Positive'' \OverwhelminglyPositive, ``Very Positive'' \VeryPositive, ``Mostly Positive'' \MostlyPositive, ``Mixed'' \Mixed, and ``Mostly Negative'' \MostlyNegative)}
 \label{tab:gamessteam}
    \resizebox{\linewidth}{!}{%
    \begin{tabular}{llll}
    \toprule
    \textbf{Video game}               & \textbf{Review}         & \textbf{Video game}                         & \textbf{Review}          \\
    \midrule
    CS:GO                             & \VeryPositive           & BeamNG drive                                & \OverwhelminglyPositive  \\
    Pubg                              & \Mixed                  & Counter strike                              & \OverwhelminglyPositive  \\
    Dota 2                            & \VeryPositive           & RimWorld                                    & \OverwhelminglyPositive  \\
    GTA V                             & \VeryPositive           & World of Tanks Blitz                        & \VeryPositive            \\
    Tom Clancy's Rainbow Six® Siege   & \VeryPositive           & The Elder Scrolls V: Skyrim Special Edition & \VeryPositive            \\
    Team fortress 2                   & \VeryPositive           & NARAKA Bladepoint                           & \MostlyPositive          \\
    Terraria                          & \OverwhelminglyPositive & Hunt: Showdown                              & \VeryPositive            \\
    Garry's Mod                       & \OverwhelminglyPositive & Civilization V                              & \OverwhelminglyPositive  \\
    Rust                              & \VeryPositive           & Project Zomboid                             & \VeryPositive            \\
    Apex                              & \VeryPositive           & Factorio                                    & \OverwhelminglyPositive  \\
    Wallpaper Engine                  & \OverwhelminglyPositive & Smite                                       & \MostlyPositive          \\
    The Witcher® 3: Wild Hunt         & \OverwhelminglyPositive & The elder scrolls online                    & \VeryPositive            \\
    Warframe                          & \VeryPositive           & theHunter: Call of the Wild™                & \VeryPositive            \\
    Destiny 2                         & \VeryPositive           & Age of Empires II: Definitive Edition       & \VeryPositive            \\
    Cyberpunk 2077                    & \MostlyPositive         & Satisfactory                                & \OverwhelminglyPositive  \\
    Dead by Daylight                  & \VeryPositive           & Stellaris                                   & \VeryPositive            \\
    ARK                               & \VeryPositive           & Fifa 22                                     & \VeryPositive            \\
    Elden ring                        & \VeryPositive           & Forza Horizon 5                             & \VeryPositive            \\
    Stardew Valley                    & \OverwhelminglyPositive & Squad                                       & \VeryPositive            \\
    Euro track simulator 2            & \OverwhelminglyPositive & The sims 4                                  & \VeryPositive            \\
    Rocket League                     & \VeryPositive           & Europa Universalis IV                       & \VeryPositive            \\
    Phasmophobia                      & \OverwhelminglyPositive & Scum                                        & \MostlyPositive          \\
    Payday 2                          & \VeryPositive           & Stumble Guys                                & \VeryPositive            \\
    The forest                        & \OverwhelminglyPositive & Assetto Corsa                               & \VeryPositive            \\
    War Thunder                       & \MostlyPositive         & Conan Exiles                                & \MostlyPositive          \\
    Valheim                           & \OverwhelminglyPositive & FINAL FANTASY XIV ONLINE                    & \VeryPositive            \\
    Brawlhalla                        & \VeryPositive           & Crusader Kings III                          & \VeryPositive            \\
    Red dead redemption 2             & \VeryPositive           & Yugioh Master Duel                          & \MostlyPositive          \\
    DayZ                              & \MostlyPositive         & Left for dead                               & \OverwhelminglyPositive  \\
    Don't Starve together             & \OverwhelminglyPositive & eFootball 2023                              & \MostlyNegative          \\
    Sea of thieves                    & \VeryPositive           & Black desert                                & \MostlyPositive          \\
    New World                         & \Mixed                  & Soundpad                                    & \OverwhelminglyPositive  \\
    Geometry Dash                     & \VeryPositive           & Total War: Warhammer 3                      & \MostlyPositive          \\
    Bloons TD 6                       & \OverwhelminglyPositive & Fallout 76                                  & \MostlyPositive          \\
    The binding of Isaac: Rebirth     & \OverwhelminglyPositive & Warhammer 40,000: Darktide                  & \Mixed                   \\
    Path of exile                     & \VeryPositive           & Moster Hunter Rise                          & \VeryPositive            \\
    Hades                             & \OverwhelminglyPositive & Coockie clicker                             & \OverwhelminglyPositive  \\
    Fallout 4                         & \VeryPositive           & EA SPORTS™ FIFA 23                          & \Mixed                   \\
    VR Chat                           & \MostlyPositive         & Farming Simulator 22                        & \VeryPositive            \\
    Lost Ark                          & \MostlyPositive         & Victoria 3                                  & \Mixed                   \\
    Civilization VI                   & \VeryPositive           & Goose Goose Duck                            & \VeryPositive            \\
    7 days to die                     & \VeryPositive           & Undecember                                  & \Mixed                   \\
    Mount  Blade II: Bannerlord       & \VeryPositive           & Mir4                                        & \Mixed                   \\
    Vampire Survivors                 & \OverwhelminglyPositive & Footbal Manager 2022                        & \VeryPositive            \\
    Cities: Skylines                  & \VeryPositive           & Dwarf fortress                              & \OverwhelminglyPositive  \\
    TmodLoader                        & \OverwhelminglyPositive & Nba 2K23                                    & \Mixed                   \\
    Arma 3                            & \VeryPositive           & Project: Playtime                           & \Mixed                   \\
    Deep rock Galactic                & \OverwhelminglyPositive & Divinity: Original Sin 2 - Definitive Edition                            &   \OverwhelminglyPositive                       \\
    Hearth of Iron IV                 & \VeryPositive           & Paragon the Overprime                       & \Mixed                   \\
    Call of Duty®: Modern Warfare® II & \Mixed                  & Football Manager 2023                       & \VeryPositive            \\
    \bottomrule
\end{tabular}
}
    \vspace{-0.3cm}
\end{table*}


\section{Experimental Procedure}
\label{sec:execution}
\begin{figure*}
 \centering\includegraphics[width=\linewidth]{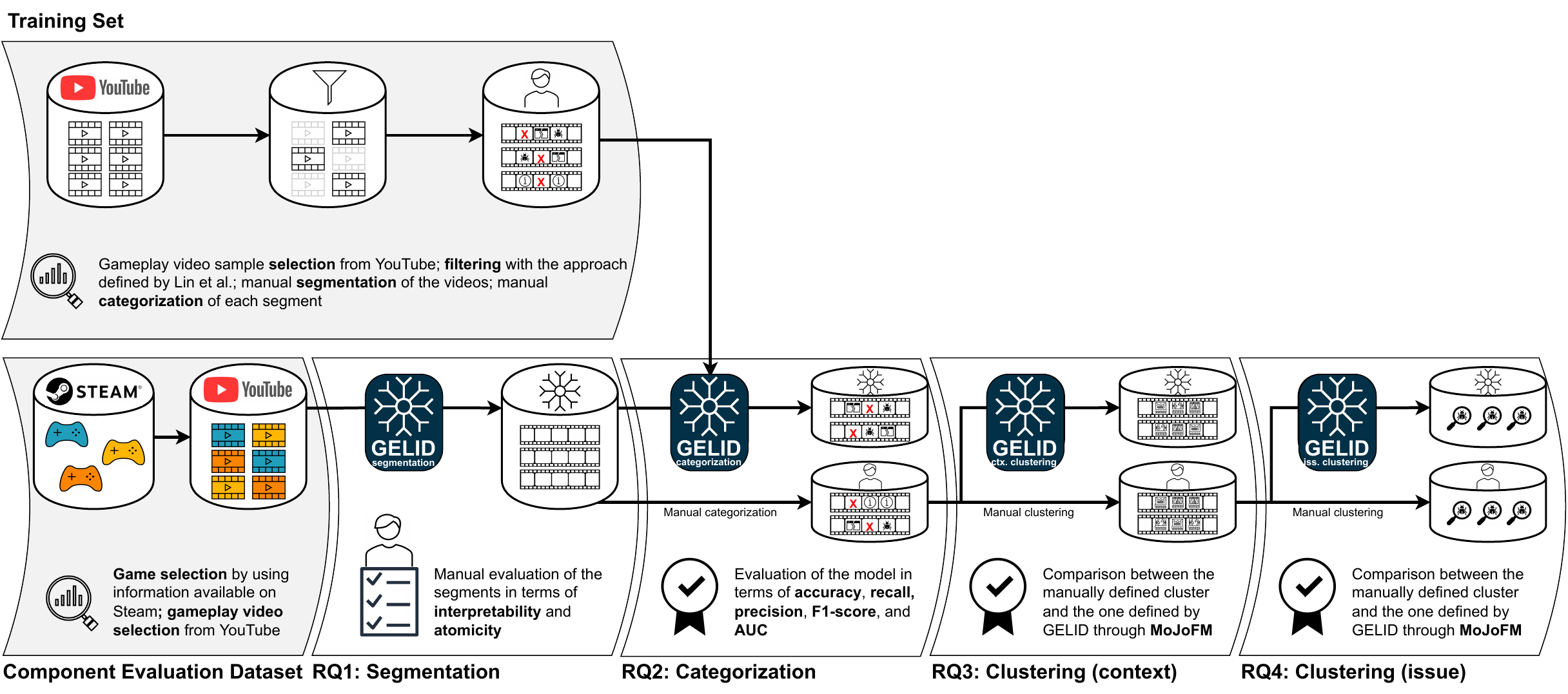}
  \captionsetup{justification=centering}
  \caption{Summary of the study design.}
  \label{fig:plan}
\end{figure*}

We summarize in \figref{fig:plan} our plan for answering the four research questions, and we provide the details below.

\subsection{Research Method for \RQ{1}: \REV{Meaningfulness of Extracted Segments}}
To answer \RQ{1}, we evaluate the technique we defined with different values of $t$ (streamer reaction times). Specifically, we instantiate our approach with $t$ in the set $\{0, 5, 10\}$ seconds. 
We ran the first step of \textit{Video Segmentation} on selected gameplay videos for each video game in the test set, collecting a total of 101 video segments. \REV{Note that the number of extracted segments is lower than the number of videos because some videos might not contain any keyword we use in the \textit{Video Segmentation} step to retrieve candidate relevant segments (see \secref{sec:videosegmentation}). Consequently, if a video does not contain any of these keywords, no segment is extracted from it.}

We evaluated the segments detected by each variant of our approach in terms of their (i) \textit{interpretability} (\ie it is possible to watch the segment and acquire all the information needed to understand what has been experienced by the streamer) (ii) the \textit{atomicity} (\ie it is not possible to further split the segments). Such aspects are complementary: It would be possible to maximize the \textit{interpretability} by creating few segments (\eg just one for the whole video); this, however, would result in lower \textit{atomicity} since the segments could be further divided into parts. \REV{While we would have ideally wanted to capture the ``quality'' of segments as a whole, it is quite hard to define a precise metric for such a complex aspect. Thus, we preferred to use two specific and easy-to-evaluate aspects instead. Concerning the relationship between such aspects and quality as a whole, we can say that, given two segments A and B, if interpretability(A) $>$ interpretability(B) and atomicity(A) $>$ atomicity(B), then quality(A) $>$ quality(B). On the other hand, if we have conflicting situations (\eg interpretability(A) $<$ interpretability(B) and atomicity(A) $>$ atomicity(B)), we can not say whether the quality of A is greater or lower than the quality of B.} 

Two of the authors watched the segments generated by each variant, for a total of 303 evaluations, and manually annotated each segment in terms of its \textit{interpretability} and \textit{atomicity} on a 5-point Likert scale. As for the first metric, we evaluated to what extent we could fully understand what is happening based only on the segment itself. As for atomicity, instead, we assessed whether the segment can be further divided in additional standalone (fully interpretable) segments. The final score was computed as 5 minus the number of additional standalone segments that could be further extracted, or 1 if more than four standalone segments were found. Each of the 303 manually analysed slices was independently inspected.
We report the inter-rater reliability between the annotators by using the Cohen's kappa coefficient \citep{cohen1960coefficient, wan2015kappa}. Then, for each segment, we compute the mean \textit{interpretability} and \textit{atomicity}.
Finally, we compare the tested techniques in terms of such metrics using a Mann-Whitney U test \citep{mann1947test, macfarland2016mann}, and adjusting the $p$-values resulting for multiple comparisons using the Benjamini and Hochberg procedure \citep{benjamini1995controlling}. 
We also report the effect size, using the Cliff's delta \citep{cliff1993dominance}, to understand the magnitude of differences observed.

\subsection{Research Method for \RQ{2}: \REV{Segment Categorization Effectiveness}}
To answer \RQ{2}, we use all the three datasets previously described. We aimed at evaluating not only the complete approach on a multi-class categorization problem (the four informative classes reported in \tabref{tab:taxonomy}, plus the \textit{non-informative} class), but also its version on a simplified version of the same problem, \ie a binary classifier (\textit{informative}, \textit{non-informative}) like the one defined by \citet{lin2019identifying}. It is worth noting, however, that we could not compare our results with the ones obtained with such an approach because it is designed to work only on entire videos, not on segments.

As a first step, we aimed at selecting (i) the best machine learning algorithm, (ii) the best set of features, and (iii) the best preprocessing pipeline for categorizing gameplay video segments in both scenarios. As candidate machine learning algorithms, we selected Random Forest \citep{ho1995random}, Logistic Regression, SMO \citep{hearst1998support}, Multilayer Perceptron \citep{ramchoun2016multilayer} and IBk\citep{choudhury2015comparative}. We used the implementations available in the Weka toolkit.\footnote{\url{http://www.cs.waikato.ac.nz/ml/weka/}} At this stage, we used the default hyperparameters available in Weka for each of them. As candidate set of features, as explained in \secref{sec:approach}, we considered three textual-based sets of features (Bag of Words, word2vec, and doc2vec), a video-based set of feature, and a mixed set of features (including both the best set of textual features and the video-based set of features).
As candidate preprocessing pipelines, we considered the use of SMOTE \citep{SMOTE}, which allows to generate synthetic instances for balancing the training set, and a two-step attribute selection approach: We first rank the features based on their respective information gain and we discard the ones with score 0; then, we run a wrapper attribute evaluator \citep{gnanambal2018classification} to select the best subset of features in terms of AUC achieved by a simple kNN model with $k=3$. More specifically, we considered four options: the use of SMOTE alone, the use of our two-step attribute selection alone, the use of both of them, and the use of none of them.
At this stage, we relied on the training set, and we performed a 10-fold cross validation for all the combinations of ML algorithms, feature sets, and preprocessing pipelines for both the problems (binary and multi-class). For each of them, we compute and report the achieved AUC (Area Under the ROC curve \citep{bradley1997use}) \citep{flach2016roc}. An AUC of 0.5 indicates a model having the same prediction accuracy of a random classifier. A perfect model (\ie zero false positives and zero false negatives) has instead AUC = 1.0. Thus, the closer the AUC to 1.0, the higher the model performances.
In the end, we select the combination that allows achieving the highest score both for the binary and the multi-class model.

Finally, as a third step, we ran the best models on the test set to understand to what extent the models would be useful in practice. In this case we report not the AUC, but also the \textit{precision}, \textit{recall}, and \textit{F-measure} scores. \textit{Precision} is computed as $\frac{\mathit{TP}}{\mathit{TP}+\mathit{FP}}$ and \textit{recall} is computed as $\frac{\mathit{TP}}{\mathit{TP}+\mathit{FN}}$, where \textit{TP}, \textit{FP}, and \textit{FN} indicate the number of \textit{true positives}, \textit{false positives}, and \textit{false negatives}, respectively. \textit{F-measure} is computed as the harmonic mean of \textit{precision} and \textit{recall}.

\subsection{Research Method for \RQ{3}: \REV{Contextual Clustering Effectiveness}}
To address \RQ{3}, we tested the two non-parametric clustering techniques described in \secref{sec:approach}, \ie DBSCAN \citep{ester1996density}, OPTICS \citep{ankerst1999optics} with two distance metric, \ie HSV and SSIM.

Both DBSCAN and OPTICS require to set an $\epsilon$ parameter, which indicates the minimum distance to be used to consider two instances belonging to the same cluster. However, determining the input parameter values can be very difficult. For both non-parametric clustering techniques, we decide the value of $\epsilon$ by using a well-known procedure \citep{ozkok2017new}. Specifically, we (i) calculate the distance between each point and its nearest neighbour, (ii) sort the distances in ascending order, (iii) compute, for each pair of consecutive distances, their difference $\Delta_i$ = $d_{i+1}$ - $d_i$), and (iv) set $\epsilon = \max(\Delta_i)$. We used this procedure independently for each clustering operation we run (\ie each combination of video game and similarity metric).

We compare the results of the algorithms with the ground-truth partition produced in the manual clustering of the test set to evaluate this step of \approach. To do this, we use the MoJo eFfectiveness Measure (\textit{MoJoFM}) \citep{wen2004effectiveness}, a normalized variant of the MoJo distance. \textit{MoJoFM} is computed using the following formula:
$$
MoJoFM(A,B)  = 100 - (\frac{mno(A,B)}{max(mno(\forall E_{A}, B))} \times 100)
$$
\noindent where $mno(A,B)$ is the minimum number of \emph{Move} or \emph{Join} operations one needs to perform in order to transform a partition $A$ into a different partition $B$, and $max(mno(\forall \; E_{A}, B)$ is the maximum possible distance of any partition $A$ from any partition $B$. \textit{MoJoFM} returns 0 if partition $A$ is the farthest partition away from $B$; it returns 100 if $A$ is equal to $B$. 

We report the MoJoFM obtained for each combination of game and metric considered.

\subsection{Research Method for \RQ{4}: \REV{Specific Issue-Based Clustering Effectiveness}}
To answer \RQ{4}, we tested the same clustering techniques considered in \RQ{3} (DBSCAN \citep{ester1996density} and OPTICS \citep{ankerst1999optics}) plus a third (\ie Mean Shift \citep{fukunaga1975estimation}) which we could not use in \RQ{3} because it can not use custom distance metrics. 
We start from the ground-truth clusters manually defined in the test set. For each of them, we run the issue-based clustering approach defined in \secref{sec:approach} on the instances belonging to them. We use the same procedure described in \RQ{3} to define the $\epsilon$ hyperparameters for DBSCAN and OPTICS for each clustering operation. This time, we do not report the $\epsilon$ values used for space reasons (given the higher number of clustering operations).
We report, like for \RQ{3}, the MoJoFM score achieved for each video game.

\subsection{Replication Package}
We publicly release in our replication package \citep{replicationpackage} the datasets used in each research question, the ARFF files used to train and test the machine learning techniques, the raw data of our manual analyses for each research question, and additional data that did not fit in our paper. We also publicly provide the implementation of each step of \approach.

	\section{Empirical Study Results}
\label{sec:results}

This section reports the results of the four research questions formulated in \secref{sec:design}.

\subsection{\RQ{1}: Interpretability and Atomicity of Gameplay Video Segments}

The IRR between the two raters when they evaluated the \textit{interpretability} of the segments extracted with \approach is $k = 0.84$, while it is $k = 0.85$ when evaluating them in terms of \textit{atomicity}. Thus, in both the cases, the agreement was \textit{almost perfect}.

\begin{figure}[t]
	\centering
	\includegraphics[width=0.49\linewidth]{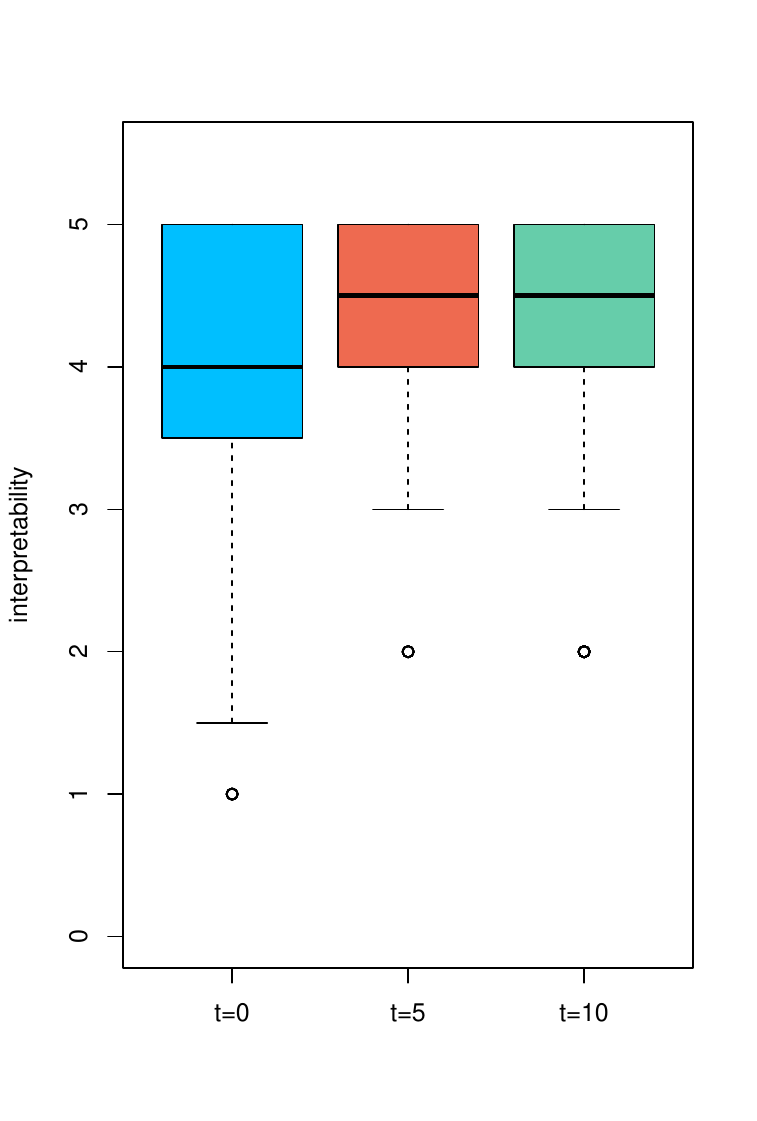}
	\includegraphics[width=0.49\linewidth]{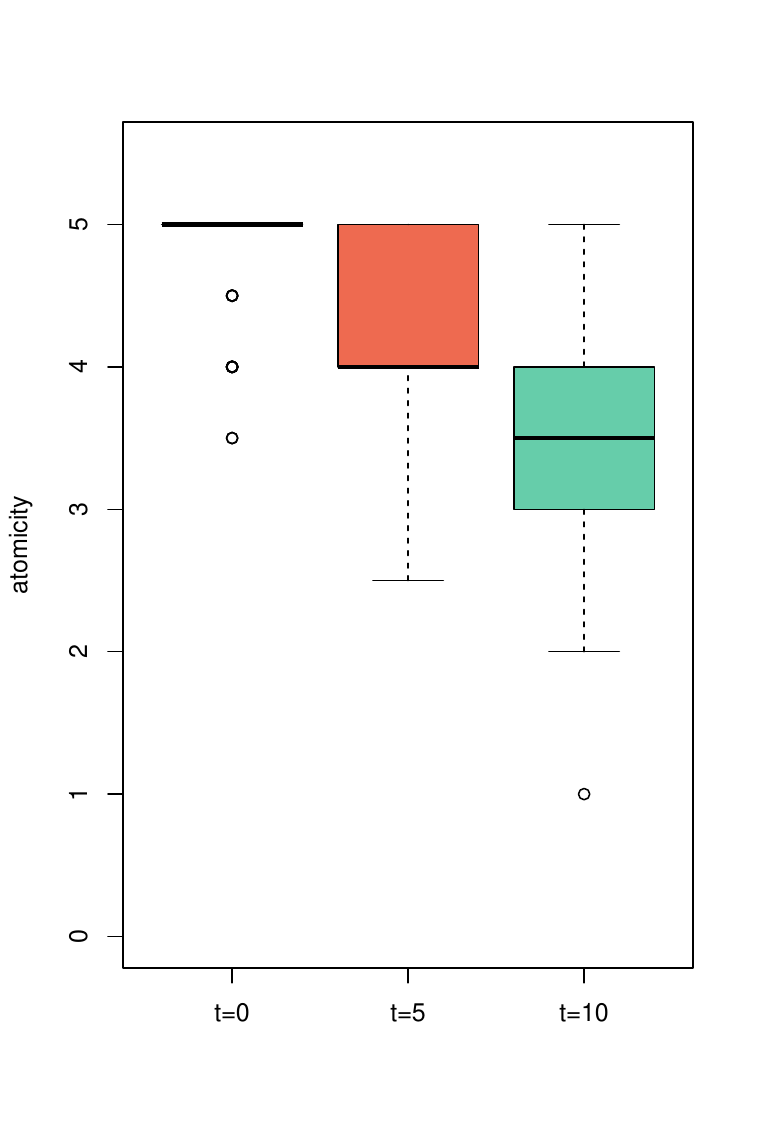}
	\caption{Distribution of interpretability (left) and atomicity (right) evaluation of gameplay video segments with the three different thresholds ($t=0$, $t=5$, $t=10$)}
	\label{fig:boxplots}
\end{figure}

When comparing $t=0$ with $t=5$ in terms of \textit{interpretability} of gameplay video segments generated by \approach, we obtain an adjusted \textit{p-value} $<$ 0.001, with a \textit{negligible} effect size ($\delta$ = -0.146). We obtain an analogous result when comparing $t=5$ with $t=10$ (p $<$ 0.001, $\delta$ = -0.092, \textit{negligible} magnitude). 
We observed also a difference between $t=0$ with $t=10$: In this case, the adjusted \textit{p-value} is the same (p $<$ 0.001), while, this time, the effect size is \textit{small} ($\delta$ = -0.227).
The boxplot in \figref{fig:boxplots} (left part) visually confirms the difference we numerically observed.

In terms of \textit{atomicity}, when comparing $t=0$ with $t=5$ we obtain an adjusted \textit{p-value} $<$ 0.001, with a \textit{large} effect size ($\delta$ = 0.610). We obtain an analogous result when comparing $t=5$ with $t=10$ (p $<$ 0.001, $\delta$ = 0.577, \textit{large} magnitude). 
As expected, again, the difference between $t=0$ with $t=10$ is \textit{large} as well (\textit{p-value} $<$ 0.001, $\delta$ = 0.869).
The boxplot in \figref{fig:boxplots} (right part) visually confirms the difference we numerically observed.
A case in which two annotators disagreed on the evaluation of the atomicity of a segment is related to segment in a gameplay video on Conan Exiles. One author rated the atomicity of the segment as 4, while the second author rated it as 5. The disagreement arose from the presence of a gameplay setting screen that appeared during the video segment, lasting about 3 seconds. This setting screen interrupted the ongoing game phase and then resumed it. The first evaluator considered this interruption significant enough to be considered as a point in which two segments could be detected, while the second annotator considered the screen appearance time negligible, given its short duration.

Considering overall the results, we can conclude that by increasing the $t$ value we obtain negligible advantages in terms of interpretability and substantial disadvantages in terms of atomicity. More specifically, while increasing $t$ from 0 to 5 allows to obtain an observable difference in terms of interpretability, having a $t$ value higher than 5 practically brings no advantage at all (see \figref{fig:boxplots} --- left part). We conclude that $t$ values higher than 5 are most likely not worth considering, while there is a trade-off that users might want to consider between $t = 0$ (which allows having substantially more atomic videos) and $t = 5$ which allows having more interpretable videos, even if slightly). Indeed, we obtain for $t = 0$ an average interpretability of 3.97 and an average atomicity of 4.88, while $t = 5$ provides an average value of 4.27 for both interpretability and atomicity.

\begin{resultbox}
 \textbf{Answer to \RQ{1}.} The proposed segmentation approach achieves satisfactory results. The best results are obtained when using $t = 0$ (privileging atomicity) and $t = 5$ (privileging interpretability).
\end{resultbox}

\subsection{\RQ{2}: Gameplay Video Segments Categorization}

\textbf{ML Pipeline Selection and Training.}
We report in \tabref{tab:rq2binary} and \tabref{tab:rq2multiclass} the results of the 10-fold cross-validation comparison performed on the training set to select the best algorithm both for binary and multi-class categorization, respectively. For deciding which sets of textual features we would include in the combination of image-based features and textual features, we compared the average results obtained with textual features alone and we picked the features that generally allow to achieve the best results (\ie Word2Vec).

The machine learning algorithm that provides the best results for binary classification is Random Forest, while the best set of features is the combination of image-based and textual features. Using both SMOTE and attribute selection, we obtained 0.79 AUC (71.8\% accuracy).
The best results could be achieved with Random Forest and a combination of image-based and textual features for multi-class categorization as well. This time, however, the best model was the one trained by only running attribute selection (\ie without balancing the training set with SMOTE). In this case, the obtained AUC is slightly lower (0.75 AUC, 62.0\% accuracy), most likely due to the inherently more difficult problem (categorizing in five classes instead of two).

\newcommand{\trotate}[1]{\begin{sideways}#1\end{sideways}}
\begin{table*}[t]
\centering
\caption{\RQ{2}: Comparison, in terms of unweighted average AUC, of different sets of features (\textbf{B}ag \textbf{o}f \textbf{W}ords, \textbf{W}ord\textbf{2}\textbf{V}ec, \textbf{D}oc\textbf{2}\textbf{V}ec, \textbf{I}mage-based features), preprocessing techniques (SMOTE and \textbf{A}ttribute \textbf{S}election), and ML algorithms for binary classification (\textit{non-informative}/\textit{informative}).}
\label{tab:rq2binary}
\begin{tabular}{l|lrrrr}
\toprule
                                              & \textbf{Model} & \textbf{Plain} & \textbf{AS} & \textbf{SMOTE} & \textbf{SMOTE + AS}  \\ \midrule
\multirow{5}{*}{\trotate{\textbf{BoW}}}       & RandomForest                       & 0.72  & 0.73                & 0.72  & 0.73                         \\
                                              & Logistic                           & 0.63  & 0.74                & 0.62  & 0.73                         \\
                                              & SMO                                & 0.68  & 0.60                & 0.68  & 0.66                         \\
                                              & MultilayerPerceptron               & 0.52  & 0.73                & 0.68  & 0.73                         \\
                                              & IBk                                & 0.58  & 0.73                & 0.60  & 0.73                         \\ \midrule
\multirow{5}{*}{\trotate{\textbf{W2V}}}       & RandomForest                       & 0.72  & 0.72                & 0.74  & 0.71                          \\
                                              & Logistic                           & 0.68  & 0.70                & 0.69  & 0.70                          \\
                                              & SMO                                & 0.65  & 0.65                & 0.65  & 0.65                          \\
                                              & MultilayerPerceptron               & 0.73  & 0.68                & 0.72  & 0.69                          \\
                                              & IBk                                & 0.62  & 0.62                & 0.62  & 0.63                          \\ \midrule
\multirow{5}{*}{\trotate{\textbf{D2V}}}       & RandomForest                       & 0.52  & 0.50                & 0.52  & 0.50                          \\
                                              & Logistic                           & 0.50  & 0.50                & 0.49  & 0.50                          \\
                                              & SMO                                & 0.51  & 0.50                & 0.48  & 0.50                          \\
                                              & MultilayerPerceptron               & 0.52  & 0.50                & 0.56  & 0.50                          \\
                                              & IBk                                & 0.50  & 0.50                & 0.52  & 0.50                          \\ \midrule
\multirow{5}{*}{\trotate{\textbf{I}}}         & RandomForest                       & 0.74  & 0.69                & 0.74  & 0.68                          \\
                                              & Logistic                           & 0.69  & 0.66                & 0.69  & 0.66                          \\
                                              & SMO                                & 0.61  & 0.58                & 0.58  & 0.57                          \\
                                              & MultilayerPerceptron               & 0.67  & 0.66                & 0.69  & 0.66                          \\
                                              & IBk                                & 0.62  & 0.61                & 0.62  & 0.60                          \\ \midrule
\multirow{5}{*}{\trotate{\textbf{W2V + I}}}   & RandomForest                       & 0.78  & 0.78                & 0.79  & \textbf{0.79}                 \\
                                              & Logistic                           & 0.70  & 0.72                & 0.70  & 0.72                          \\
                                              & SMO                                & 0.68  & 0.65                & 0.67  & 0.63                          \\
                                              & MultilayerPerceptron               & 0.74  & 0.73                & 0.76  & 0.71                          \\
                                              & IBk                                & 0.65  & 0.66                & 0.64  & 0.66                          \\
\bottomrule
\end{tabular}
\end{table*}

\begin{table*}[t]
\centering
\caption{\RQ{2}: Comparison, in terms of unweighted average AUC, of different sets of features (\textbf{B}ag \textbf{o}f \textbf{W}ords, \textbf{W}ord\textbf{2}\textbf{V}ec, \textbf{D}oc\textbf{2}\textbf{V}ec, \textbf{I}mage-based features), preprocessing techniques (SMOTE and \textbf{A}ttribute \textbf{S}election), and ML algorithms for multi-class classification (\textit{logic}, \textit{presentation}, \textit{performance}, \textit{balance}, \textit{non-informative}).}
\label{tab:rq2multiclass}
\begin{tabular}{llrrrr}
\toprule
                                              & \textbf{Model} & \textbf{Plain} & \textbf{AS} & \textbf{SMOTE} & \textbf{SMOTE + AS}  \\ \midrule
\multirow{5}{*}{\trotate{\textbf{BoW}}}       & RandomForest                       & 0.72  & 0.70                 & 0.71   & 0.69                          \\
                                              & Logistic                           & 0.70  & 0.69                 & 0.70   & 0.70                          \\
                                              & SMO                                & 0.67  & 0.62                 & 0.67   & 0.69                          \\
                                              & MultilayerPerceptron               & 0.70  & 0.71                 & 0.70   & 0.70                          \\
                                              & IBk                                & 0.60  & 0.69                 & 0.60   & 0.69                          \\ \midrule
\multirow{5}{*}{\trotate{\textbf{W2V}}}       & RandomForest                       & 0.73  & 0.72                 & 0.70   & 0.71                          \\
                                              & Logistic                           & 0.59  & 0.71                 & 0.63   & 0.69                          \\
                                              & SMO                                & 0.67  & 0.60                 & 0.70   & 0.67                          \\
                                              & MultilayerPerceptron               & 0.67  & 0.64                 & 0.68   & 0.65                          \\
                                              & IBk                                & 0.58  & 0.59                 & 0.61   & 0.60                          \\ \midrule
\multirow{5}{*}{\trotate{\textbf{D2V}}}       & RandomForest                       & 0.52  & 0.49                 & 0.51   & 0.49                           \\
                                              & Logistic                           & 0.48  & 0.49                 & 0.48   & 0.49                          \\
                                              & SMO                                & 0.49  & 0.50                 & 0.49   & 0.49                          \\
                                              & MultilayerPerceptron               & 0.52  & 0.49                 & 0.50   & 0.49                          \\
                                              & IBk                                & 0.49  & 0.49                 & 0.51   & 0.49                          \\ \midrule
\multirow{5}{*}{\trotate{\textbf{I}}}         & RandomForest                       & 0.69  & 0.62                 & 0.67   & 0.62                          \\
                                              & Logistic                           & 0.66  & 0.64                 & 0.65   & 0.64                          \\
                                              & SMO                                & 0.54  & 0.53                 & 0.62   & 0.61                          \\
                                              & MultilayerPerceptron               & 0.52  & 0.66                 & 0.63   & 0.64                          \\
                                              & IBk                                & 0.56  & 0.58                 & 0.56   & 0.56                          \\ \midrule
\multirow{5}{*}{\trotate{\textbf{W2V + I}}}   & RandomForest                       & 0.74  & \textbf{0.75}        & 0.74   & 0.71                          \\
                                              & Logistic                           & 0.61  & 0.71                 & 0.65   & 0.70                          \\
                                              & SMO                                & 0.70  & 0.57                 & 0.71   & 0.68                          \\
                                              & MultilayerPerceptron               & 0.71  & 0.67                 & 0.71   & 0.66                          \\
                                              & IBk                                & 0.60  & 0.59                 & 0.62   & 0.58                          \\
\bottomrule
\end{tabular}
\end{table*}

\textbf{Testing the Models.}
\tabref{tab:rq2binary_videogames} and \tabref{tab:rq2multiclass_videogames} report the recall, precision, F-Measure and AUC
scores achieved by the best model for binary and multi-class categorization, respectively. In detail, we report the results achieved both for individual games and for all the instances together. 

Overall, the binary classification model exhibits slightly worse results compared to the ones obtained on the training set with 10-fold cross validation (0.61 AUC vs. 0.79). The model has an acceptable recall (72\%) and a relatively low precision (56\%) on the \textit{informative} class. This means that a developer would be able to get most of the potentially interesting segments, but they also have to manually discard many non-informative ones in the process. The results, however, depend much on the video game at hand: For Conan Exiles, for example, the model always achieves acceptable results both in terms of overall precision (66\%) and recall (64\%). This might depend on many factors. First, on the quality of the streaming videos taken into account: Streamers might be more verbose for some video game genres, thus allowing the classifier to better identify the segments. Second, on the similarity with video games included in the training set: Some genre- or game-specific terms might be indicative of an issue for some games, while not for others. For example, the phrase ``loot hack'' might appear in online multiplayer role play games and indicate a \textit{logic} issue, but it might not be pronounced at all by streamers playing racing games.

Analogous conclusions can be drawn from the results achieved with the multi-class model. In this case, it is interesting to observe that some classes the classifier never categorizes instances as \textit{performance} and \textit{balance} (``\textit{--}'' for precision in \tabref{tab:rq2multiclass_videogames}). This is possibly due to the fact that such issue types are generally less prevalent than others\footnote{33 and 48 in the training set, 31 and 4 in the test set for \textit{performance} and \textit{balance}, respectively.} and, thus, the model fails to learn how to recognize them. It is also worth noting that we were not able to find \textit{balance} issues in one of the games taken into account (\ie DayZ). Overall, the model achieves better results on the \textit{presentation} class. This is probably due to the fact that, for this category, the model also relies on image-based features, which are less relevant for the other classes.

\begin{table*}[t]

\newcommand{\I}{\faIcon[regular]{info}}
\newcommand{\NI}{\faIcon[regular]{trash}}
\newcommand{\OV}{\faIcon[regular]{globe}}
\centering
\caption{\RQ{2}: Performance of the best binary categorization model on the test set. We use the icon \I{} to indicate the \textit{informative} class and the icon \NI{} to indicate the \textit{non-informative} class, while \OV{} indicates their weighted mean.}
\label{tab:rq2binary_videogames}
\resizebox{\linewidth}{!}{%
\begin{tabular}{l|rrr|rrr|rrr|rrr}
\toprule
\multirow{2}{*}{\textbf{Game}}     & \multicolumn{3}{c|}{\textbf{Precision}} & \multicolumn{3}{c|}{\textbf{Recall}} & \multicolumn{3}{c|}{\textbf{F-Measure}} & \multicolumn{3}{c}{\textbf{AUC}} \\
                                   & \I    & \NI    & \OV                     & \I    & \NI    & \OV                 & \I    & \NI    & \OV                    & \I    & \NI    & \OV             \\
\midrule                                                                                                                                                                                         
{Conan Exiles}                     & 61\% & 70\%  & 66\%                      & 77\% & 52\%  & 64\%                  & 68\% & 60\% & 64\%                     & 0.73  & 0.73 & 0.73              \\
{DayZ}                             & 52\% & 54\%  & 53\%                      & 71\% & 34\%  & 53\%                  & 60\% & 42\% & 51\%                     & 0.55  & 0.55 & 0.55              \\
{New World}                        & 65\% & 47\%  & 58\%                      & 71\% & 40\%  & 59\%                  & 68\% & 43\% & 59\%                     & 0.61  & 0.61 & 0.61              \\
\midrule                                                                                                                                                                                                
{Overall}                          & 56\% & 62\%  & 60\%                      & 72\% & 45\%  & 58\%                  & 63\% & 52\% & 58\%                     & 0.58  & 0.64 & 0.61              \\
\bottomrule
\end{tabular}
}
\end{table*}

\begin{table*} [t]
\newcommand{\Lo}{\faIcon[regular]{bug}}
\newcommand{\PR}{\faIcon[regular]{ghost}}
\newcommand{\Pf}{\faIcon[regular]{clock}}
\newcommand{\Bl}{\faIcon[regular]{balance-scale-right}}
\newcommand{\NI}{\faIcon[regular]{trash}}
\newcommand{\OV}{\faIcon[regular]{globe}}
\centering
\caption{\RQ{2}: Performance of the best multi-class categorization model on the test set. We use the icons \Lo{}, \PR{}, \Pf{}, \Bl, and \NI{} to indicate the \textit{logic}, \textit{presentation}, \textit{performance}, \textit{balance}, and \textit{non-informative} classes, respectively, while \OV{} indicates their weighted mean.}
\label{tab:rq2multiclass_videogames}
\resizebox{\linewidth}{!}{%

\begin{tabular}{l|rrrrrr|rrrrrr}
\toprule                                   
\multirow{2}{*}{\textbf{Game}}     & \multicolumn{6}{c|}{\textbf{Precision}}                   & \multicolumn{6}{c}{\textbf{Recall}} \\
                                   & \NI    & \PR    & \Lo   & \Bl   & \Pf   & \OV             & \NI    & \PR    & \Lo   & \Bl   & \Pf   & \OV  \\
\midrule                                                                                                                                                                                         
\textbf{Conan Exiles}              & 58\%   & 48\%   & 25\%  & -     & -     & 48\%               & 81\%   & 39\%   &  3\%  & 0\%   & 0\%   & 55\%    \\
\textbf{DayZ}                      & 51\%   & 35\%   & 0\%   & -     & -     & 39\%               & 61\%   & 34\%   &  0\%  & -     & 0\%   & 43\%    \\
\textbf{New World}                 & 56\%   & 49\%   & 25\%  & -     & -     & 48\%               & 71\%   & 40\%   & 50\%  & 0\%   & 0\%   & 52\%    \\
\midrule                                                                                                                                                                                                                 
\textbf{Overall}                   & 56\%   & 44\%   & 13\%  & -     & -     & 45\%               & 73\%   & 38\%   & 10\%  & 0\%   & 0\%   & 51\%    \\
\bottomrule

\toprule
\multirow{2}{*}{\textbf{Game}}     & \multicolumn{6}{c|}{\textbf{F-Measure}}                   & \multicolumn{6}{c}{\textbf{AUC}} \\
                                   & \NI    & \PR    & \Lo   & \Bl   & \Pf   & \OV             & \NI    & \PR    & \Lo   & \Bl   & \Pf   & \OV  \\
\midrule                                   
\textbf{Conan Exiles}              & 68\%   & 43\%   &  5\%  & -     & -     & 49\%               & 0.69   & 0.65   & 0.58  & 0.33  & 0.57  & 65\%    \\
\textbf{DayZ}                      & 55\%   & 35\%   &  0\%  & -     & -     & 43\%               & 0.52   & 0.52   & 0.53  & -     & 0.47  & 51\%    \\
\textbf{New World}                 & 63\%   & 44\%   & 33\%  & -     & -     & 49\%               & 0.64   & 0.60   & 0.93  & 0.72  & 0.10  & 62\%    \\
\midrule                                                                                                                                              
\textbf{Overall}                   & 63\%   & 40\%   & 10\%  & -     & -     & 47\%              & 0.63   & 0.60   & 0.54  & 0.61  & 0.48  & 60\%    \\
\bottomrule
\end{tabular}
}
\end{table*}

\begin{resultbox}
 \textbf{Answer to \RQ{2}.} The categorization models defined are not able achieve satisfactory results both for binary and multi-class categorization.
\end{resultbox}

\subsection{\RQ{3}: Clustering Gameplay Video Segments by Context}

\begin{figure}[t]
	\centering
	\includegraphics[width=0.49\linewidth]{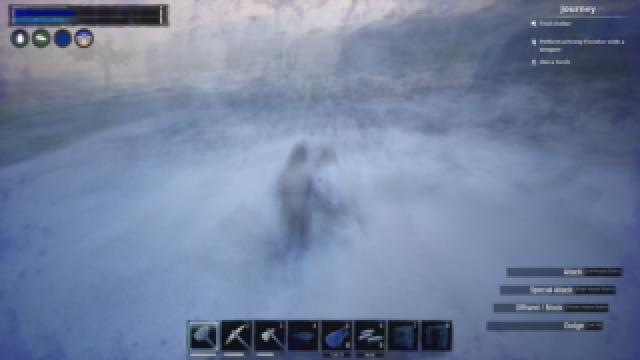}
	\includegraphics[width=0.49\linewidth]{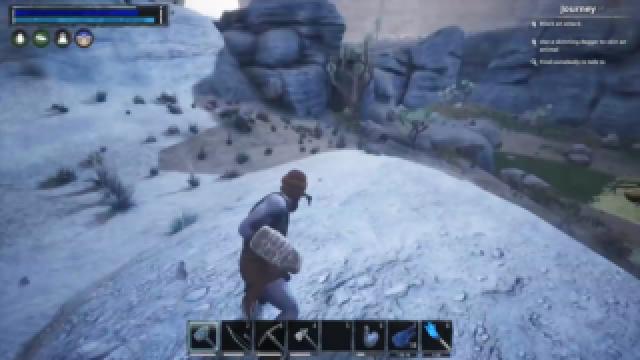}
	\caption{Different game scenes in Conan Exiles grouped in the same context cluster}
	\label{fig:clusterconan}
\end{figure}

\tabref{tab:context_clustering} shows the MoJoFM score achieved by the two tested algorithms when comparing their output with the manually defined clusters. First, it can be observed that OPTICS allows to achieve the best results for all games taken into account, between 46.0\% (New World) and 21.9\% (Conan Exiles). It is worth noting that the variability among video games is, in this case, quite high. This is expected: Some games have areas and levels very similar one to another, thus making the task of visually distinguishing the areas quite challenging even for a human who never played the game.
For example, the frames presented in \figref{fig:clusterconan} represent two visually similar areas in Conan Exiles that, however, are different.

Overall, however, we can conclude that the clustering approach we defined in \approach is only partially able to achieve its goal.

\begin{table}[h]
\centering
\caption{\RQ{3}: MoJoFM achived for clustering by context with HSV}
\label{tab:context_clustering}
\begin{tabular}{l r r }
\toprule
                             & \textbf{DBSCAN}  & \textbf{OPTICS} \\
\midrule
\textbf{Conan Exiles}        & 17.8\%           & 21.9\%      \\
\textbf{DayZ}                & 23.2\%           & 36.6\%        \\
\textbf{New World}           & 28.0\%           & 46.0\%       \\
\textbf{Average}             & 23.0\%           & 34.8\%     \\
\bottomrule
\end{tabular}
\end{table}

\begin{table}[h]
\centering
\caption{\RQ{3}: MoJoFM achived for clustering by context with SSIM}
\label{tab:context_clustering}
\begin{tabular}{l r r}
\toprule
                             & \textbf{DBSCAN}  & \textbf{OPTICS}  \\
\midrule
\textbf{Conan Exiles}        & 4.8\%           & 21.9\%       \\
\textbf{DayZ}                & 0.0\%           & 2.5\%        \\
\textbf{New World}           & 0.0\%           & 18.5\%       \\
\textbf{Average}             & 1.6\%           & 14.3\%        \\
\bottomrule
\end{tabular}
\end{table}

\begin{resultbox}
 \textbf{Answer to \RQ{3}.} We obtained mixed results for the clustering by context step because its performance strongly depends on the video game at hand.
\end{resultbox}

\subsection{\RQ{4}: Clustering Gameplay Video Segments by Specific Issues}
\tabref{tab:issue_clustering} shows the MoJoFM score achieved by the three tested algorithms when comparing their output with the manually defined clusters. 
In this case, the results are definitely better than the ones obtained in the previous experiment, with the best-performing algorithm (DBSCAN) achieving 72.7\% MoJoFM score. This is due to the fact that, in this case, there were less instances to cluster for two of the games taken into account (DayZ and New World). As a result, the task was inherently easier. It is worth noting, however, that for Conan Exiles the number of instances to cluster was quite large, in some cases, up to 18 and DBSCAN still achieves very good results (71.2\% MoJoFM).

Differently from what observed for \RQ{3}, we have a much less marked variance among the games (between 69.1\% and 77.8\%). OPTICS, in this case, achieved slightly worse results than DBSCAN, while MeanShift is clearly less effective than the others.

\begin{table} [h]
\centering
\caption{\RQ{4}: MoJoFM achived for clustering on the specific-issue.}
\label{tab:issue_clustering}
\begin{tabular}{lrrrr}
\toprule
                             & \textbf{DBSCAN}  & \textbf{OPTICS}    & \textbf{MeanShift} \\
\midrule                                                             
\textbf{Conan Exiles}        & 71.2\%           & 62.5\%             & 52.9\%             \\
\textbf{DayZ}                & 69.1\%           & 69.1\%             & 58.2\%             \\
\textbf{New World}           & 77.8\%           & 77.8\%             & 55.6\%             \\
\textbf{Average}             & 72.7\%           & 69.8\%             & 55.5\%             \\
\bottomrule
\end{tabular}
\end{table}

\begin{resultbox}
 \textbf{Answer to \RQ{4}.} DBSCAN allows to cluster the segments very similarly to how human annotators clustered them, with a low variability among video games.
\end{resultbox}

	\section{Discussion}
\label{sec:discussion}

The main problems we encountered are in the automated categorization of issues in gameplay video segments and in the context-based segment clustering (steps 2 and 3 of \approach).  

First, it is worth noting that our results partially contrast the ones obtained by \citet{lin2019identifying}, who defined a classifier able to correctly distinguish \textit{informative} from \textit{non-informative} gameplay videos. Segment-level categorization is a much harder problem than video-level categorization. This is confirmed by the fact that even simplifying our five-class categorization problem in binary categorization problem (similarly to the one addressed by \citet{lin2019identifying}, but on segments), we still obtain negative results (58\% F-Measure, with 0.61 AUC). We have some hypothesis on why this is the case. First, videos have metadata (such as tags, descriptions, and so on) that segments lack. \citet{lin2019identifying} used such metadata, but we could not use them in our context. If a video is specifically aimed at reporting issues (\ie it contains a compilation of game errors), it is very likely that the authors explicitly mention this in the description. Gameplay video subtitles, instead, are much more noisy. 

We observed that, often, the subtitle sentences are incomplete and ambiguous (\eg ``logics, bro. Well, I talk all'' used for a \textit{logic} problem, ``they are lower than that'' used for a \textit{presentation} problem, and ``less well-known logic that's arguably one'' used for a \textit{performance} problem). To some extent, this happens because the comment corresponding to the portion of the video in which the issue appears might not be in sync with the issue itself: Gamers might talk about the issues even several minutes after it appears. It is worth noting that this problem is not related to the automated segmentation, because in evaluating step 2 with \RQ{2} we used manually-defined segments. The problem is in the lack of (logical) sync between what streamers say and when what they say happens on screen.
Future work could consider a larger context for extracting the features (\eg the surrounding $n$ seconds, with even large values of $n$) instead of only considering the subtitles related to the specific segment. \REV{The idea based on the possibility of using a larger context stems from the assumption that expanding the context of observation allows for a broader view of what is happening in the specific gameplay video, thus in the game, and allows more features to be extracted.}

\begin{resultbox}
 \textbf{Lesson Learned 1.} Considering a larger context for extracting textual features might allow obtaining better results.
 
 \noindent \REV{\textbf{Future Research Idea 1.}  To overcome this limitation, future research could aim to consider a larger portion of video both before and after the given identified segment.}
\end{resultbox}

\REV{Using keywords to detect possibly useful segments of the gemeplay videos might be detrimental. Indeed, there may be segments without streamer comments, that would be completely ignored. These are blind spots for GELID. To address this limitation, it may be necessary to develop new and specialized approaches to detect specific problems, such as glitches or stuttering events.}

Related to this, another problem we noticed by analyzing some examples is that streamers sometimes comment on their gaming experience in an irregular manner, often even through simple exclamations (\eg ``the glitch myself?'' for \textit{performance}, ``BAM!'' for \textit{logic}, ``and there!'' for \textit{presentation}).
Catching those issues is probably infeasible by only relying on textual information. Similarly, we can observe a performance problem found in a gameplay video of New World:\footnote{\url{https://youtu.be/1duizy5DSOg?t=1540}} The game temporarily freezes while the player is running, but they say \textit{``here can see one right now okay stop doing that let's start running they're nasty big aren't they''}, referring to what is happening in the game. Automatically categorizing this kind of issue is, again, extremely challenging, and a more specific approach would be needed.
\REV{Another limitation of GELID is related to the fact that it only relies on gameplay videos in English. Future work is needed to generalize it to other languages. In CLAP \citep{scalabrino2017listening}, an attempt has been made to deal with this issue. The authors tried to translate the input textual information (in the context of GELID, subtitles) from foreign languages into English and then use the normal approach (which works on English) to deal with them. However, this solution proved to be unsuccessful. In this paper, we use word2vec: It would be possible to test the effectiveness of word2vec models trained on other languages. Based on the negative results obtained for English, which is quite widespread, we believe that the implementation of such an approach cannot be successful at present.}
\begin{resultbox}
 \textbf{Lesson Learned 2.} Sometimes, textual features are not useful at all since the streamers use generic exclamations to report issues. 
 
 \noindent \REV{\textbf{Future Research Idea 2.} Future research could aim at taking into account the slang used by streamers and to define a vocabulary of the terms most commonly used to describe different kinds of issues or to define specialized approaches to detect issues mostly based on the videos rather than on the captioned spoken content.}
\end{resultbox}

When looking at the multi-class categorization, the problem is even more evident in terms of general effectiveness of the model. We report in \tabref{tab:multiclass_confusion_matrix} the confusion matrix for the multi-class categorization model. While the model correctly identifies 81 \textit{presentation} issues, it correctly detects only 2 \textit{logic}-related issues and, again, no \textit{performance}- and \textit{balance}-related issue. More interestingly, the model often categorizes \textit{presentation}-related issues as \textit{logic} issues, while the opposite happens relatively less frequently. In general, instead, the model tends to confuse the specific categories of instances as \textit{presentation}-related, probably because it is the most frequent informative type of issue.

We analyzed some misclassified instances, aiming at getting some insights on why the model tends to confuse some \textit{presentation} issues for \textit{logic} issues and why it is not able to correctly identify \textit{performance} and \textit{balance} instances.
We found an interesting example in DayZ. The streamer says \textit{``my doesn't seem to be archived it back back is so annoying''},\footnote{\url{https://youtu.be/eDQIdqDC-sc?t=239}} but the model probably confuses the indication of an ``annoying'' circumstance for something related to a functional issue (\textit{logic}), while, in this case, it was referred to a \textit{presentation} issue.

\begin{resultbox}
 \textbf{Lesson Learned 3.} Given the strong class unbalance, categorization does not work well for detecting \textit{performance} and \textit{balance} problems. Approaches specifically designed for finding such categories of issues might be needed.
 
 \noindent \REV{\textbf{Future Research Idea 3.} To increase the number of \textit{balance} and \textit{performance} instances, it could be useful to look for and specifically take into account video games that are or have been notorious for such problems.} 
\end{resultbox}

\begin{table*}[h]
\newcommand{\Lo}{\faIcon[regular]{bug}}
\newcommand{\PR}{\faIcon[regular]{ghost}}
\newcommand{\Pf}{\faIcon[regular]{clock}}
\newcommand{\Bl}{\faIcon[regular]{balance-scale-right}}
\newcommand{\NI}{\faIcon[regular]{trash}}
\centering
\caption{\RQ{2}: Confusion Matrix for multi-class categorization on all the instances (Conan Exiles, DayZ, and New World). The columns indicate the categories assigned by the classifier, while the rows indicate actual ones.}
\label{tab:multiclass_confusion_matrix}
\begin{tabular}{l|rrrrr}
         & \NI             & \PR          & \Lo        & \Pf         & \Bl         \\ 
\midrule                                                   
\NI      & \textbf{218}    & 75           & 5          & 0           & 0           \\
\PR      & 126             & \textbf{81}  & 8          & 0           & 0           \\
\Lo      & 27              & 16           & \textbf{2} & 0           & 0           \\
\Pf      & 18              & 13           & 0          & \textbf{0}  & 0           \\
\Bl      & 3               & 1            & 0          & 0           & \textbf{0}  \\
\bottomrule
\end{tabular}
\end{table*}

Another possible reason behind the failure in categorization could be related to the procedure used to define the training set: To collect an adequate number of instances, we considered videos that explicitly report issues (\ie that contain keywords such as ``bug'' in their title or description). It is possible that these videos are intrinsically different from the long gameplay videos we used for testing the models. To check if this is the case, we trained/tested two classifiers (both for binary and multi-class categorization) based on the best configurations found in \RQ{2} by using 10-fold cross validation on the test set alone, both globally and by considering the instances of single games. We report the results in \tabref{tab:testastraining}. We observed a clear increase in the effectiveness of both the models, with the binary classification model achieving $\sim$82\% accuracy on two games. While more data would be necessary, the results of this analysis suggest that videos explicitly reporting issues are too different from long gameplay videos (that we aim to target) in which issues sometimes appear. Thus, it would be more appropriate to build the training set using the same procedure used to build the test set, even if this require a much bigger effort (it would not be possible, for example, to use the approach by \citet{lin2019identifying} as a filter). Also, using a training set composed of only game-specific instances might allow to achieve better results (even if we observed this only for two games out of three). \REV{In detail, again, a training set defined on a specific game allows for more precise information in relation to the game area/level. For example, open world games have very similar game areas, so a large amount of data would allow a more precise distinction to be made between the different game areas in which users find themselves.}

\begin{resultbox}
 \textbf{Lesson Learned 4.} A training set built on long gameplay videos not specifically aimed at reporting issues might help achieving better results. Also, game-specific training might help increasing the model accuracy.
  
 \noindent \REV{\textbf{Future Research Idea 4.} Future research should verify what is the impact of the type of video, \ie long and generic gameplay videos or short and focused gameplay videos reporting issues, on the performance of the four steps of GELID.}
\end{resultbox}

\begin{table}
 \caption{Accuracy and AUC achieved by training/testing the best models for binary and multi-class categorization on the test set alone using 10-fold cross validation.}
 \label{tab:testastraining}
 \centering
 \begin{tabular}{l|rr|rr}
\toprule
\multirow{2}{*}{\textbf{Game}}     & \multicolumn{2}{c|}{\textbf{Binary}} & \multicolumn{2}{c}{\textbf{Multi-class}} \\
                                   & Accuracy    & AUC        & Accuracy & AUC  \\
\midrule
{Conan Exiles}                     & 81.7\%      & 0.89       & 71.7\%   & 0.73 \\
{DayZ}                             & 64.7\%      & 0.75       & 59.0\%   & 0.56 \\
{New World}                        & 81.7\%      & 0.89       & 67.9\%   & 0.72 \\
\midrule                                                                                                                                                                                                
{Combined}                         & 72.7\%      & 0.79       & 59.9\%   & 0.63 \\
\bottomrule
\end{tabular}
\end{table}

As for the context-based segment categorization (step 3 of \approach), as we previously mentioned while analyzing the results, the poor performance can be due to the fact that some games have visually similar, but logically different game areas/levels. Some video games might suffer from this issue more than others. In our case, we observed that our approach  (specifically, the variant based on HSV histogram correlation, which achieves the best results) works reasonably well on New World, but remarkably bad on Conan Exiles. For the video games on which our approach does not work well, a more sophisticated (and game-specific) approach might be used, which should be specialized on the game at hand so that, for example, it is able to distinguish the specific game areas by recognizing specific game elements.

\begin{resultbox}
 \textbf{Lesson Learned 5.} A game-specific approach for recognizing the game area/level might be needed for some video games.
 
 \noindent \REV{\textbf{Future Research Idea 5.} Researchers should test the impact of introducing game- or game-genre-specific features on the effectiveness of the context-based clustering.}
\end{resultbox}

	\section{Threats to Validity}
\label{sec:threats}

\textbf{Threats to construct validity} mainly pertain the possible imprecisions made while defining the test set used to evaluate \approach and to answer all our research questions. As explained in \secref{sec:design}, to reduce this threat, two evaluators independently tagged each instance and discussed conflicts aiming at reaching consensus. This occurred in 1.2\% of the cases.

\textbf{Threats to internal validity} concern factors internal to our study that could have affected the results. 
A first threat regarding \RQ{2} is related to the specific set of ML techniques we decided to use and to the preprocessing pipelines we tested. As for the first, we took into account the main categories of classic ML approaches. It is possible that Deep Learning-based approach achieve better results, but we avoided using such approaches because even a small Neural Network (Multilayer Perceptron) achieves very poor results given the small size of our training set. \REV{Another limitation related to \RQ{2} is the choice not to tune the hyperparameters and to use the predefined hyperparameters provided by Weka. To understand the impact of this decision, we tried to replicate the results of binary classification of segments as \textit{informative} and \textit{non-informative} while varying the main hyperparameter for Random Forest (\ie the maximum number of features). We report in \tabref{tab:rq2_hyperparameters} the results of such an analysis on the Conan Exiles dataset.\footnote{Note that the results differ from the ones reported in \tabref{tab:rq2binary_videogames} because we did not run any preprocessing step here.} Although this analysis revealed some improvements in model performance while varying such a parameter, we found that the impact of not tuning it was rather small (+4 percentage points for F-Measure and +0.01 for AUC). Thus, we believe this is not the cause of the negative results we obtained.}

\begin{table*}[t]

\newcommand{\I}{\faIcon[regular]{info}}
\newcommand{\NI}{\faIcon[regular]{trash}}
\newcommand{\OV}{\faIcon[regular]{globe}}
\centering
\caption{\REV{\RQ{2}: Hyperparameter Tuning of the Random Forest categorization model on Conan Exiles. We use the icon \I{} to indicate the \textit{informative} class and the icon \NI{} to indicate the \textit{non-informative} class, while \OV{} indicates their weighted mean.}}
\label{tab:rq2_hyperparameters}
\resizebox{\linewidth}{!}{%
\begin{tabular}{l|rrr|rrr|rrr|rrr}
\toprule
\multirow{2}{*}{\textbf{NumFeatures}}     & \multicolumn{3}{c|}{\textbf{Precision}} & \multicolumn{3}{c|}{\textbf{Recall}} & \multicolumn{3}{c|}{\textbf{F-Measure}} & \multicolumn{3}{c}{\textbf{AUC}} \\
                                   & \I    & \NI    & \OV                     & \I    & \NI    & \OV                 & \I    & \NI    & \OV                    & \I    & \NI    & \OV             \\
\midrule                                                                                                                                                                                         
{Unlimited (default)}                    & 55\% & 72\%  & 63\%                      & 88\% & 29\%  & 59\%                  & 68\% & 42\% & 55\%                     & 0.68  & 0.68 & 0.68              \\
{1}                             & 55\% & 65\%  & 60\%                       & 80\% & 37\% & 58\%                 & 65\% & 47\% & 56\%                     & 0.61  & 0.61 & 0.61              \\
{2}                        & 56\% & 74\%  & 65\%                      & 87\% & 34\%  & 60\%                  & 68\% & 47\% & 58\%                     & 0.63  & 0.63 & 0.63              \\
{3}                     & 57\% & 80\%  & 69\%                      & 92\% & 32\%  & 62\%                  & 70\% & 45\% & 57\%                     & 0.66  & 0.66 & 0.66              \\
{4}                             & 54\% & 66\%  & 60\%                      & 85\% & 27\%  & 56\%                  & 66\% & 39\% & 52\%                     & 0.62  & 0.62 & 0.62              \\
{5}                        & 56\% & 79\%  & 68\%                      & 92\% & 31\%  & 61\%                  & 70\% & 45\% & 57\%                     & 0.65  & 0.65 & 0.65              \\
{6}                     & 55\% & 80\%  & 68\%                      & 93\% & 26\%  & 59\%                  & 70\% & 39\% & 54\%                     & 0.69  & 0.69 & 0.69              \\
{7}                             & 56\% & 77\%  & 69\%                      & 70\% & 45\%  & 57\%                  & 69\% & 45\% & 57\%                     & 0.67  & 0.67 & 0.67              \\
{8}                        & 65\% & 47\%  & 58\%                      & 56\% & 79\%  & 68\%                  & 92\% & 31\% & 61\%                     & 0.69  & 0.69 & 0.69              \\
{9}                     & 55\% & 72\%  & 63\%                      & 88\% & 30\%  & 59\%                  & 68\% & 42\% & 55\%                     & 0.69  & 0.69 & 0.69              \\
{10}                             & 57\% & 84\%  & 70\%                      & 93\% & 32\%  & 63\%                  & 71\% & 47\% & 59\%                     & 0.68  & 0.68 & 0.68              \\
\bottomrule
\end{tabular}
}
\end{table*}

The classes we consider for the multi-class categorization problem (\RQ{2} might be incomplete: It is possible that we do not consider some relevant categories of issues. To mitigate this threat, we avoided defining such categories based on our personal experience, but we relied on a state-of-the-art taxonomy \cite{truelove2021we}.
A key threat regards the features considered for step 2 (and, thus, to answer \RQ{2}). It is worth noting that we relied on features that proved to be useful in other contexts (\eg categorization and clustering of mobile app reviews \cite{chen2014ar, scalabrino2017listening}), and we also augmented them with video-based features. Still, it is possible that a different set of features leads to better results.
As for clustering (both \RQ{3} and \RQ{4}), it is possible that we chose sub-optimal parameters (\ie $\epsilon$ values). To reduce this threat, we used a rigorous procedure \cite{ozkok2017new} to set these values for each tested video game.

Finally, \textbf{threats to external validity} concern the generalizability of our findings. Our test set is composed of gameplay videos related to only 3 video games. We could not select videos from a more diverse set of video games because we needed multiple segments related to the same game areas to address \RQ{3} and \RQ{4}. However, it is worth noting that we also report in \tabref{tab:rq2binary} and \tabref{tab:rq2multiclass} the results of a 10-fold cross validation performed on the training set, which, differently from the test set, is composed of videos from many video games 110, specifically. Nevertheless, we acknowledge that most of our results are not necessarily generalizable to the vast quantity of video game genres and video games available in the market. 

\REV{We believe that the variety of video games is not as relevant as the variety and type of streamers involved. GELID heavily relies on (captioned) spoken content for segmentation and categorization. To this end, having verbose streamers could benefit GELID. On the other hand, the video game selection might mostly impact the two clustering-related steps: For example, games with many graphically similar levels or areas might deceive GELID while it cluster segments.}

	\section{Conclusion}
\label{sec:conclusions}
In recent years, there has been a growing interest in video games. During game development, many bugs go undetected prior to release because of the difficulty of fully testing all aspects of a video game.
We introduce \approach, a novel approach for detecting anomalies in video games from gameplay videos to support developers by providing them with useful information on how to improve their games. We validated the single steps of \approach in an empirical study involving 604 segments extracted from 80 hours of gameplay videos related to 3 video games (Conan Exiles, DayZ, and New World). We obtained mixed results: The effectiveness of both segmentation (step 1) and issue-based clustering (step 4) are satisfactory, while we observed that categorization (step 2) and context-based clustering (step 3) of segments still do not work sufficiently well to be used in practice.
Future work should aim at addressing these two problems. To foster research in this field, we publicly release all the (manually annotated) datasets in our replication package \cite{replicationpackage}.


	\section{Data Availability Statement}
	All the datasets produced and the scripts implemented to obtain the results reported in this paper (including our implementation of \approach) are available in our replication package \citep{replicationpackage}.


	\section*{Conflict of interest}
	
	The authors declare that they have no conflict of interest.

	
	\balance
	\bibliographystyle{spbasic}
	\bibliography{main}
	
\end{document}